\documentclass[12pt,preprint]{aastex}
\usepackage{epsfig,amssymb}
\usepackage{graphicx}
\usepackage{rotating}
\usepackage{mathrsfs}
\usepackage{natbib}
\newcommand{\ben}{\begin{enumerate}}
\newcommand{\een}{\end{enumerate}}
\newcommand{\bit}{\begin{itemize}}
\newcommand{\eit}{\end{itemize}}
\newcommand{\bfig}{\begin{figure}}
\newcommand{\efig}{\end{figure}}
\newcommand{\beq}{\begin{equation}}
\newcommand{\eeq}{\end{equation}}
\newcommand{\mbf}{\mathbf}

\newcommand{\mcal}{\mathcal}

\shorttitle{Active-Region Energetics and Helicity II}
\shortauthors{Georgoulis, Tziotziou, \& Raouafi}

\begin{document}
\title{Magnetic Energy and Helicity Budgets in the Active-Region Solar
  Corona. II. Nonlinear Force-Free Approximation}


\author{Manolis K. Georgoulis{\footnote{Marie Curie Fellow}} \& Kostas Tziotziou}
\affil{Research Center for Astronomy and Applied Mathematics (RCAAM)\\
       Academy of Athens, 4 Soranou Efesiou Street, Athens, Greece, GR-11527}
\and 
\author{Nour-Eddine Raouafi}
\affil{The Johns Hopkins University Applied Physics Laboratory (JHU/APL)\\
       11100 Johns Hopkins Rd. Laurel, MD 20723-6099, USA}
\altaffiltext{1}{Marie Curie Fellow.}
\begin{abstract}
Expanding on an earlier work that relied on linear force-free magnetic fields, we self-consistently derive the instantaneous free magnetic energy and relative magnetic helicity budgets of an unknown three-dimensional nonlinear force-free magnetic structure extending above a single, known lower-boundary magnetic field vector. The proposed method does not rely on the detailed knowledge of the three-dimensional field configuration but is general enough to employ only a magnetic connectivity matrix on the lower boundary. The calculation yields a minimum free magnetic energy and a relative magnetic helicity consistent with this free magnetic energy. The method is directly applicable to photospheric or chromospheric vector magnetograms of solar active regions. Upon validation, it basically reproduces magnetic energies and helicities obtained by well-known, but computationally more intensive and non-unique, methods relying on the extrapolated three-dimensional magnetic field vector. We apply the method to three active regions, 
calculating the photospheric connectivity matrices by means of simulated annealing, rather than a model-dependent nonlinear force-free extrapolation. For two of these regions we correct for the inherent linear force-free overestimation in free energy and relative helicity that is larger for larger, more eruptive, active regions. In the third studied  region, our calculation can lead to a physical interpretation of observed eruptive manifestations. We conclude that the proposed method, including the proposed inference of the magnetic connectivity matrix, is practical enough to contribute to a physical interpretation of the dynamical evolution of solar active regions. 
\end{abstract}
\section{Introduction}
\label{S-Intro}
Several decades have elapsed since the notion of magnetic helicity and its application to solar magnetic fields were introduced. This considerable time has been marked by an impressive volume of published works on the subject. Yet, we are still lagging behind in understanding, first, how to practically calculate magnetic helicity in the Sun and, second, what is the actual role of magnetic helicity in solar eruptions. 

Undoubtedly, the apparent lack of a breakthrough in this topic stems from our crucially incomplete knowledge of solar magnetic fields: unable to observe their generation at the solar interior and to precisely measure them in the solar atmosphere, we can routinely detect and measure them only in the photospheric and/or low-chromospheric interface. To calculate magnetic helicity, however, we need either the three-dimensional magnetic field in all or part of the solar coronal volume, including its lower boundary 
\citep{woltjer_58, berger_84, finn_antonsen_85, berger_99}
or the flow velocity field on this lower boundary 
\citep{berger_field_84, chae_01, demoulin_berger_03}. 
Both defy a unique calculation, showing critical model-dependent ambiguities either when  extrapolating for the coronal magnetic field (see the comprehensive comparisons of \citet{schrijver_etal_06}
and \citet{metcalf_etal_08}) or when inferring a reliable photospheric velocity field 
\citep{november_simon_88, kusano_etal_02, nindos_zhang_02, nindos_etal_03, longcope_04, schuck_05, schuck_08, georgoulis_labonte_06, welsch_etal_07, ravindra_etal_08, chae_sakurai_08}. 
Knowledge of the three-dimensional coronal magnetic field of, say, an active region, is needed to calculate the instantaneous magnetic helicity (and energy) budgets in the region, while knowledge of the local flow field at the lower atmospheric boundary is necessary to calculate the injection rate of magnetic helicity in the solar atmosphere due to the region's temporal evolution. The accumulated helicity budgets are then obtained by integrating the helicity injection rate in time \citep[e.g.,][]{green_etal_02, nindos_etal_03, labonte_etal_07}.

When the role of magnetic helicity in solar eruptions is examined, there are studies  suggesting that helicity is not necessary for flares and coronal mass ejections (CMEs) 
\citep{phillips_etal_05, zuccarello_etal_09}. At the same time, other studies suggest that helical (or helical by proxy) active regions tend to be the most eruptive ones 
\citep{nindos_andrews_04, labonte_etal_07, nindos_09, georgoulis_etal_09}.
One should acknowledge the fact that solar eruptions can occur, at least in models, in the absence of significant magnetic helicity accumulations. However, since helicity is a signed quantity with right-handed (positive) and left-handed (negative) senses, absence of a significant helicity budget could also mean significant helicity accumulation of {\it both} senses at roughly similar amounts -- this can lead even to the so-called ``helicity annihilation'' that is a proposed eruption mechanism \citep{kusano_etal_03}. Moreover, several eruption mechanisms stem from instabilities that do not explicitly rely on helicity. Such mechanisms are the magnetic flux cancellation 
\citep{vanball_martens_89}; ``hoop'' force \citep{chen_96}; breakout \citep{antiochos_etal_99}; tether-cutting \citep{moore_etal_01}; and the torus instability \citep{kliem_torok_06}, among others. On the other hand, a popular eruption mechanism that relies on magnetic helicity is the helical kink instability 
\citep{rust_kumar_96, baty_etal_98, kliem_torok_05, kliem_etal_12}. Moreover, it has been shown that injection of helicity in a modeled eruption results in faster CMEs after a helicity threshold is exceeded \citep{zuccarello_etal_09}.

In theory, therefore, solar eruptions can occur with or without a significant magnetic helicity budget, namely without a {\it dominant} helicity sense, although eruptions may be necessary to diffuse into the heliosphere the excess helicity produced in the Sun \citep{low_94, rust_94, rust_03}. 
This is because helicity is known to be roughly invariant in a volume enclosing an isolated magnetic structure even under resistive manifestations such as magnetic reconnection (e.g., \citet{berger_99} and references therein). On the other hand, virtually all eruption mechanisms, regardless of helicity dependence, result in strongly helical eruption ejecta widely known as flux ropes. As to the pre-eruption situation, we cannot collect clues about the actual role played by helicity unless we (i) compare the helicity budgets between non-eruptive and eruptive active regions in the pre-eruption state, and (ii) assess the relevance of proposed eruption initiation mechanisms with observations in a very detailed manner. 

A prerequisite of both objectives above is the practical and reliable calculation of magnetic helicity in observed solar magnetic structures. A first step in this direction was taken by \citet{georgoulis_labonte_07} -- hereafter Paper I. In Paper I we described a methodology to simultaneously calculate the relative magnetic helicity and the free magnetic energy, with respect to a potential-field reference, of a magnetic structure represented by a single solar vector magnetogram. The underlying assumption was the validity of the linear force-free (LFF) field approximation in the magnetic structure. Calculation of both the relative helicity and the free energy was physically consistent and did not rely on a prescribed flow velocity field or the detailed three-dimensional coronal field above the lower-boundary magnetogram. In essence, it was a convenient {\it surface} calculation of physical quantities that stem from the three-dimensional magnetic field on and above the surface. The main drawback of the 
methodology, however, was its central assumption of a constant-$\alpha$, LFF magnetic structure -- an assumption that is known to be unrealistic in both the solar surface and the overlying corona \citep{metcalf_etal_95, georgoulis_labonte_04, socas-navarro_05}. Nonetheless, comparison between a non-eruptive and an eruptive solar active region revealed, even beyond large uncertainties inherent to the LFF field approximation, that the most profound differences between the two regions occurred in their budgets of free energy and relative helicity: for a factor of $\sim 3$-difference in unsigned magnetic flux between the two regions the energies and relative helicities were different by a factor of $\sim 9$, with the largest values assigned to the eruptive active region. As we show in this work this very large difference is partly due to the adopted LFF field approximation.   

In Paper I we explicitly stated that the proposed methodology would serve as the basis for a more realistic, nonlinear force-free (NLFF) field approximation in calculating the magnetic energy and relative magnetic helicity. We take this step in this work. The analysis of Paper I is extended to derive the {\it self} terms of free energy and relative helicity, while we draw from the study of 
\citet{demoulin_etal_06} to derive the {\it mutual} terms of these quantities. By construction, the LFF field methodology of Paper I treated a given magnetic structure as a single, isolated, force-free flux tube and hence it was unable to predict mutual energy and helicity terms occurring due to the interaction between different flux tubes. This work assumes a collection of discrete, slender force-free flux tubes with variable force-free parameters and hence calculates both self and mutual terms of energy and helicity. As in Paper I, this NLFF field approach is a surface calculation that does not use three-dimensional field extrapolations or velocity fields. Instead, the proposed method uses a {\it magnetic connectivity matrix} on the boundary where the vector magnetogram is obtained. This matrix can be obtained in any way possible, be it a field extrapolation or not. Therefore, our method is general and applies to any connectivity matrix, regardless of inference. To provide perspective, we apply the method to the same active-
region magnetograms as in Paper I and compare the results. 

The study is structured as follows: the methodology of the calculation is given in Section \ref{S-method}. The adopted validation procedure and its results are given in Section \ref{S-val}. Section \ref{S-res} provides the numerical results obtained by applying the method to three different solar active regions. Section \ref{S-fin} discusses our findings  and provides our conclusion and future perspective. 
\section{Methodology}
\label{S-method}
\subsection{Magnetic connectivity matrix and $\alpha$-distribution}
\label{S-method1}
The first task is to translate a continuous vector magnetogram into a
collection of discrete force-free flux tubes with known footpoints, flux contents, and
different force-free parameters $\alpha$. 
If the three-dimensional coronal magnetic
field configuration was available, then one would be able to trace each
magnetic field line separately (here the footpoint of 
a ``field line'' is restricted to the resolution element [pixel] of the studied magnetogram). The coronal configuration may be assessed by
extrapolations of various sophistication levels (i.e., current-free, LFF- or
NLFF-field) but the true configuration is unknown. 
Moreover, tracing and analyzing each field line 
separately would be impractical and unnecessary. For this reason we 
simplify the vector magnetogram into a collection of thin 
flux tubes as follows: 
\ben
\item[1.] We translate the magnetic field configuration into an
  ensemble of ``magnetic charges'' using the flux partition method
  introduced in the magnetic charge topology model of \citet{barnes_etal_05}.
  This is a flux tessellation scheme that relies on a modified   
  downhill-gradient minimization algorithm with certain provisions about saddle   
  points. This step requires only the normal (vertical) magnetic field
  component $B_z$. The chosen thresholds for partitioning a magnetogram for
  this work are (i) a threshold of $50\;G$ in $|B_z|$, (ii) a minimum magnetic flux of
  $10^{20}\;Mx$ per partition, and (iii) a minimum area of 40 magnetogram
  pixels per partition. These criteria are set to prevent the inclusion of quiet-Sun, weak-field, and very small-scale structures, respectively, into the calculation, unnecessarily adding to both complexity and required computing time. Only partitions that satisfy all three threshold criteria are selected for further analysis. Upon completion, we 
  can readily assess the flux content and flux-weighted centroid position
  of each magnetic flux partition.   
\item[2.] Assuming that flux partitioning returned $p$ positive-polarity and $n$
  negative-polarity magnetic partitions, together with their respective fluxes 
  $\mathscr{F}_i$; $i \equiv \{ 1,...,p \}$ and 
  $\mathscr{F}_j$; $j \equiv \{ 1,...,n \}$, one may define a $p \times n$
  magnetic-flux connectivity matrix. The matrix will contain the fluxes 
  $\mathscr{F}^{con}_{ij}$ committed to the connection $ij$ between the 
  $i$-positive-polarity and the $j$-negative-polarity partition. Obviously, 
  $\mathscr{F}^{con}_{ij} = 0$ in case the two partitions are not connected. Along 
  with the flux connectivity matrix we construct one more $p \times n$ matrix 
  containing the vector positions of the two flux-weighted
  centroids of connected partitions. 
\item[3.] Each magnetic connection is hereafter 
  assumed a slender flux tube with flux
  content $\mathscr{F}^{con}_{ij}$ and footpoints corresponding to the
  flux-weighted centroids of the two involved partitions. To determine the
  force-free parameter $\alpha$ of this tube we find the
  $\alpha$-parameters for each partition. From the force-free approximation
  one may easily deduce that the flux-weighted mean $\alpha$-value over a magnetic
  partition of flux $\mathscr{F}$ is given by 
  \beq
  \alpha = {{4 \pi} \over {c}} {{I} \over {\mathscr{F}}}\;\;,
  \label{a_part}
  \eeq
  where $I$ is the total electric current of the partition and $c$ is the speed of light. 
  The total current $I$ can be calculated by using the integral form of Amp\'{e}re's law 
  on the lower boundary magnetic-field vector $\mbf{B}$, i.e. 
  \beq 
  I = \oint _{\mathscr{C}} \mbf{B} \cdot d \mbf{l}\;\;,
  \label{Itot}
  \eeq
  where integration occurs along the closed contour $\mathscr{C}$ surrounding
  the partition. 

  On the practical side, a valid question is how to determine the 
  bounding contour 
  $\mathscr{C}$ of the partition in order to evaluate Equations
  (\ref{a_part}), (\ref{Itot}). 
  The partition shapes cannot be modeled easily since a
  partition can assume any closed-curve shape without restriction. 
  To determine 
  $\mathscr{C}$ and its contiguous order 
  of points we have developed an ``edge
  tracker'' that minimizes the length of the curve bounding the
  partition. Minimization is performed by iteratively choosing pairs of 
  neighboring boundary points. This is a classical optimization problem that we 
  solve iteratively via a simulated annealing method \citep{press_etal_92}.

  Let $\alpha_i$, $\alpha_j$ be the calculated force-free parameters of the
  two partitions $i$ and $j$. We assign a force-free parameter 
  \beq
  \alpha _{ij} = {{1} \over {2}} (\alpha_i + \alpha_j)\;\;,  
  \label{a_ij}
  \eeq
  for the resulting connection. For each of the two $\alpha$-values $\alpha_i$,  
  $\alpha_j$ there are respective   
  uncertainties $\delta \alpha_i$ $\delta \alpha_j$ due to the 
  uncertainties $\delta I$ in the calculation of the total current $I$ (Equation \ref{a_part}), assuming that the magnetic flux $\mathscr{F}$ is known without uncertainty. 
  The respective uncertainty 
  $\delta \alpha _{ij}$ is, then, 
  \beq
  \delta \alpha _{ij} = {{1} \over {2}} \sqrt{\delta \alpha_i^2 + \delta   
  \alpha_j^2}\;\;. 
  \label{da_ij}
  \eeq    
\een

The flux connectivity matrix $\mathscr{F}^{con}_{ij}$ described, we now discuss how
we populate it. Obviously, the result of any magnetic field extrapolation can
be translated into a connectivity matrix by tracing all extrapolated field
lines that open and close within the lower boundary. At this stage, we ignore magnetic connections closing beyond the limits of the finite lower boundary. 
Tracing closed field lines from
footpoint to footpoint, we add the flux contents of field lines that are
rooted in the same pair of partitions, thus constructing 
$\mathscr{F}^{con}_{ij}$. The simplest connectivity matrix,
$\mathscr{F}^{con}_{ij_{pot}}$, is the one obtained by a current-free (potential)
field extrapolation \citep[e.g.,][]{schmidt_64, alissandrakis_81}.
Any non-potential extrapolation can
also be used here, but if we use a NLFF field extrapolation we will reach a
non-unique result subject to the details of the extrapolation
method. For this reason, our method of choice is the simulated annealing method introduced by \citet{georgoulis_rust_07}. The method minimizes the
magnetic flux imbalance simultaneously with the separation length (footpoint
distance) of the chosen flux tubes thus emphasizing connections between
tightly arranged ensembles of 
flux partitions, most notably in active regions with
pronounced magnetic polarity inversion lines. We have revised the original
concept of \citet{georgoulis_rust_07} to (i) include a mirror flux
distribution with as much positive- and negative-polarity magnetic flux as the
negative- and positive-polarity magnetic flux of the original magnetogram at
large (more than twice the diagonal length of the original magnetogram) distances, thus
producing an exactly flux-balanced magnetic structure and treating large-scale, ``open'' magnetic connections, and (ii) include a constant normalization length $R_{max}$ 
equal to the largest length scale of the enlarged, 
flux-balanced magnetogram. These revisions result in a {\it unique}
connectivity matrix $\mathscr{F}^{con}_{ij}$ for the chosen minimization functional 
\beq
M = \sum _{ij}( {{|\mbf{r}_i - \mbf{r}_j|} \over {R_{max}}} + 
               {{|\mathscr{F}_{i}+\mathscr{F}_{j}|} \over 
                {|\mathscr{F}_{i}|+|\mathscr{F}_{j}|}} )\;\;.
\label{Mfunc}
\eeq

The above revisions to the original simulated annealing scheme of \citet{georgoulis_rust_07} alleviate the criticism applied by \citet{barnes_leka_08}. First, these authors argued that the connectivity result of \citet{georgoulis_rust_07} was not unique, depending on the origin of the coordinate system used, because $R_{max}$ in Equation (\ref{Mfunc}) was originally $|\mbf{r}_i| + |\mbf{r}_j|$. Although tests with different system origins showed little, if any, impact for the resulting connectivity, the introduction of the fixed $R_{max}$ puts this issue to rest. Moreover, \citet{barnes_leka_08} claimed that simulated annealing yields an unphysical connectivity matrix that matches neither the potential-field connectivity nor the true coronal connectivity. Due to our inability to measure the three-dimensional magnetic field vector in the corona, however, the "true" connectivity is unknown. Therefore, one cannot comment on its similarity, or difference thereof, with the connectivity revealed by simulated 
annealing. We continue to rely on annealing because it emphasizes connectivity in tightly organized active regions, that are statistically the most eruptive ones. Point taken, the methodology discussed here is more general and can accommodate {\it any} connectivity matrix. 

In Figure \ref{mtx_fig} we show an example of connectivity calculation in NOAA active region (AR) 10254, recorded by the Imaging Vector Magnetograph
(IVM; \citealt{mickey_etal_96, labonte_etal_99}) on 2003 January 13.
The difference between the potential-field and the simulated-annealing connectivity is obvious with the latter committing more flux to fewer, more closely seated, partitions. Also shown is the map of the flux-weighted 
$\alpha$-value of each flux partition (Equation (\ref{a_part})). 
\subsection{Magnetic Energy and Relative Magnetic Helicity Budgets in the NLFF
  Field Approximation}
\label{S-method2}
Consider a set of $N$ discrete magnetic flux tubes in force-free equilibrium. The magnetic helicity of this set can be viewed as the sum of all terms present in a diagonal matrix $N \times N$. Diagonal terms $l=m$ ($l,m \equiv \{1,...,N \}$) 
correspond to {\it self-helicity} terms and are due to the helical features of each flux tube independently. Off-diagonal terms $l \ne m$ are due to the interaction between pairs $(l,m)$ of flux tubes and correspond to {\it mutual-helicity} terms. For an {\it open} volume, where the set of flux tubes permeates a lower boundary and extends in the half space above it, \citet{demoulin_etal_06} showed that the relative (with respect to that of a potential field) magnetic helicity of the set can be written as 
\beq
H_m = \sum _{l=1}^N T_l \Phi_l^2 + \sum _{l=1}^N  \sum _{m=1, l \ne m}^N 
      \mcal{L}_{lm} \Phi _l \Phi _m\;\;.
\label{Hm_gen}
\eeq
The two terms of the rhs of Equation (\ref{Hm_gen}) correspond to the self and mutual helicity of the set of flux tubes, respectively. Here $T_l$ is the self-helicity factor of the flux tube $l$ with a magnetic flux content $\Phi_l$ and is due to its internal structure (twist and writhe). Furthermore, $\mcal{L}_{lm}$ is the mutual-helicity factor due to the interaction of a given pair $(l,m)$ of different flux tubes. For the studied open volume, \citet{demoulin_etal_06} further found   
\beq
\mcal{L}_{lm} = \mcal{L}_{lm}^{close} + \mcal{L}_{lm}^{arch}\;\;,
\label{L_ij}
\eeq
where $\mcal{L}_{lm}^{close}$ is the Gauss linking number, a signed integer reflecting the number and sense of the turns 
a flux tube $l$ winds around a flux tube $m$ and vice versa (see also \citet{moffatt_ricca_92}), and $\mcal{L}_{lm}^{arch}$ is the mutual-helicity factor of two arch-like flux tubes that are not winding around each other. This factor is a real number and can be attributed to the translational motions needed to bring the tubes from infinity to their prescribed footpoint positions. Derivation of various $\mcal{L}_{lm}^{arch}$-values was  described by \citet{demoulin_etal_06} and is further explained in Appendix A, where additional cases pertinent to our analysis appear. 

Further assuming {\it slender} flux tubes, i.e. flux tubes with typical diameter that is much smaller than their footpoint separation, to make use of the outcome of the connectivity matrix of Section \ref{S-method1}, the mutual-helicity term of Equation (\ref{Hm_gen}) describes the mutual helicity of the set calculated by the flux-tube axes \citep{demoulin_etal_06}. However, both $T_l$ and $\mcal{L}_{lm}$ are unknown unless knowledge of the sub-photospheric closures and coronal shapes of the flux tubes is available. 

After defining the relative magnetic helicity of the set of flux tubes, it is necessary 
to define the corresponding free magnetic energy $E_c$. \citet{demoulin_etal_06} provided an expression for $E_c$ only in the case of a {\it closed} volume, where the entire length of the closed flux tubes is known and visible. This expression cannot be used here. This being said, Paper I and \citet{berger_88} defined an {\it energy-helicity} formula in the NLFF field approximation. The free magnetic energy in this formula reads 
\beq
E_c = {{1} \over {8 \pi}} \int _\mcal{V} \alpha \mbf{A} \cdot \mbf{B} dV\;\;,
\label{Ec_or}
\eeq
where $\mcal{V}$ is the integration volume. Obviously $\mbf{A} \cdot \mbf{B}$, where $\mbf{A}$ is the vector potential ($\nabla \times \mbf{A} = \mbf{B}$), reflects the volume density of the relative magnetic helicity provided that $\mbf{A}$ obeys the Coulomb gauge (Paper I). In the simplified, constant-$\alpha$ (LFF field) approximation, Equation (\ref{Ec_or}) reduces to $E_c = \alpha / (8 \pi) H_m$. From Equation (\ref{Ec_or}), but also from the necessity to ensure that 
$E_c \rightarrow 0$ when $H_c \rightarrow 0$, we approximate the free magnetic energy in our set of $N$ discrete, slender flux tubes by the expression 
\beq
E_c = {{1} \over {8 \pi}} 
      \sum _{l=1}^N \alpha_l T_l \Phi_l^2 + 
      {{1} \over {8 \pi}} \sum _{l=1}^N  \sum _{m=1, l \ne m}^N 
      \alpha_l \mcal{L}_{lm} \Phi _l \Phi _m\;\;,
\label{Ec_gen}
\eeq
thus employing different force-free parameters $\alpha _l$ for different flux tubes $l$. Besides ensuring self-consistency, Equation (\ref{Ec_gen}) makes sure that 
$E_c \rightarrow 0$ {\it faster} than $\alpha$, because in case of a potential magnetic field, $T_l$ tends to zero and $\mcal{L}_{lm} \Phi _l \Phi _m$-terms algebraically sum to zero, at least in a flux-balanced magnetic structure (see Section \ref{S-special}). That {\it both} $E_c$ and $H_m$ must tend to zero for potential fields is also a necessity: to view this simply, 
consider the LFF field approximation again where $H_m \sim E_c / \alpha$. In case of nearly potential fields, where $\alpha \rightarrow 0$, if $E_c$ tends to zero slower than $\alpha$ then $|H_m| \rightarrow \infty$. We have shown in Paper I that indeed 
$H_m \rightarrow 0$ when $\alpha \rightarrow 0$. 

Its advantages given, a weakness of Equation (\ref{Ec_gen}) is that it is qualitatively similar to the current-channel description of \citet{melrose_04}. \citet{demoulin_etal_06} argued convincingly that this description is not equivalent to the flux-tube description attempted here. This is because each existing flux tube $l$ ($l \equiv \{1,...,N \}$) should spawn a number of additional potential flux tubes beyond it in a space-filling, force-free configuration. These additional flux tubes induce additional terms in Equation (\ref{Ec_gen}) when interacting with the non-potential flux tubes of the set. Otherwise put, for Equation (\ref{Ec_gen}) to be valid, the field should not be space-filling, that is, "sheath'' currents should contain each of the $N$ flux tubes that become then embedded in a field-free space occupied by non-magnetized plasma. This might be valid in the photosphere but is not the case in the corona \citep[e.g.,][]{longcope_welsch_00, georgoulis_etal_12}, where fields are thought to be force-free. Therefore, Equation (\ref{Ec_gen}) is not fully consistent with the NLFF field approximation that we pursue in this study.

Despite shortcomings, however, Equation (\ref{Ec_gen}) can still serve as a {\it lower limit} of the magnetic free energy. This is because the additional terms induced by the spawned potential flux tubes should only add positive increments to $E_c$. Validating our results in Section \ref{S-val}, we show that $E_c$ from Equation (\ref{Ec_gen}) is indeed a {\it realistic} lower limit for the magnetic free energy of the NLFF field.

The description of magnetic energy budgets of the set of flux tubes is complete when the reference (potential) energy $E_p$ is calculated. This can be done in more than one ways, by simply using the normal field component $B_z$ at the anchoring, boundary plane of the flux tubes. In particular, using the Virial theorem
\citep{molodensky_74, aly_84} in its frequently-used form (see Paper I for a
discussion), we have  
\beq
E_p = {{1} \over {4 \pi}} \int _{\mcal{S}} \mbf{r} \cdot \mbf{B_p} B_z
d \mcal{S}\;\;,
\label{Ep_vir}
\eeq
where $\mbf{r}$ is a vector position with arbitrary origin in the
anchoring plane $\mcal{S}$ and $\mbf{B_p}$ is
the potential-field vector on $\mcal{S}$. Moreover, by means of the expression 
\beq
E_p = {{1} \over {4 \pi}} \int _{\mcal{S}} \mbf{B_p} \times \mbf{A_p}
\cdot \hat{z} d \mcal{S}\;\; 
\label{Ep_PI}
\eeq
derived in Paper I, where $\nabla \times \mbf{A_p} = \mbf{B_p}$ and
$\hat{z}$ is the upward-pointing unit vector normal to the plane
$\mcal{S}$. For a reliable calculation of $E_p$ on $\mcal{S}$
one needs a magnetic structure as flux-balanced as possible. Finally, by 
volume-integrating the energy density $B_p^2/(8 \pi)$ of the potential field $\mbf{B_p}$. In Paper I all methods were shown to provide nearly identical $E_p$-values. Then, the total magnetic energy $E_t$ of the studied magnetic structure becomes 
\beq
E_t = E_p + E_c\;\;.
\label{E_tot}
\eeq

In conclusion, the common, self-consistent definition of $H_m$ and $E_c$ will allow us to derive both quantities from observed solar vector magnetograms without requiring sub-photospheric or coronal information. This practical way of calculating NLFF field free-energy and relative-helicity budgets is described in the following. 
\subsubsection{Self terms}
\label{S-method2a}
The self terms of the magnetic free energy and the relative magnetic
helicity refer exclusively to the twist and writhe of each
magnetic flux tube independently and are given by 
\beq
E_{c_{self}} ={{1} \over {8 \pi}} \sum _{l=1}^N \alpha_l T_l^{close} \Phi _l^2
\;\;\;and\;\;\;
H_{m_{self}} = \sum _{l=1}^N T_l \Phi_l^2\;\;, 
\label{self1}
\eeq
respectively. Both terms will be calculated, not by the above formulas, but by 
generalizing the linear analysis of Paper I. 

For the free energy of a 
single force-free flux tube we derived in Paper I the linearized
expression  
\beq
E_c = \mcal{F}_{lin} d^2 \alpha ^2 E_p\;\;,
\label{Ec_lin} 
\eeq
where $\alpha$ is the unique force-free parameter, $d$ is the linear
size element (the pixel size, in case of observed magnetograms),
and $\mcal{F}_{lin}$ is a linearized scale factor calculated in Fourier
space. Assuming a collection of force-free flux tubes rather than a single tube,
Equation (\ref{Ec_lin}) describes the self term of the free energy under
the condition that $\alpha$ is fixed, i.e. the same for all tubes. For
different $\alpha$-values per tube, that is, in case of the NLFF field
approximation, Equation (\ref{Ec_lin}) would need to be evaluated for
{\it each} tube. This is untenable, however, 
because the potential energy $E_p$ (Equations (\ref{Ep_vir}),
(\ref{Ep_PI})) and $\mcal{F}_{lin}$ are calculated once, for the
entire plane $\mcal{S}$ of the magnetogram and not for each flux
tube separately. To address this problem, thus generalizing the analysis of
Paper I, we investigate the relation between the ``scaled'' potential
energy $\mcal{F}_{lin} E_p$ and the magnetic flux $\Phi$, in case of a
single flux tube, or a magnetogram envisioned as a single flux
tube. For a large number of magnetograms this relation is
shown in Figure \ref{scat_plot}. The straight 
line represents the least-squares best fit and reveals a robust power-law
scaling of the form   
\beq
\mcal{F}_{lin} E_p = A \Phi ^{2 \lambda}\;\;,
\label{srel}
\eeq
where $A = 10^{-16.731 \pm 0.08}$ and $\lambda = 1.153 \pm 0.002$ are
the scaling constants. The substantial statistics of Figure
\ref{scat_plot} stem from the fact that, to calculate
$\mcal{F}_{lin}$, $E_p$, and $\Phi$, one does not need vector
magnetograms; therefore we have used tens of thousands of
active-region magnetograms recorded by the Michelson Doppler Imager
(MDI; \citealt{scherrer_etal_95}) onboard the Solar and Heliospheric
Observatory (SoHO) mission. All these magnetograms were located within 
$\pm 30^o$ from the central solar meridian; hence, the line-of-sight field
corresponds to the vertical field component within
15\%. The correction of \citet{berger_lites_03} was applied
to compensate for the underestimation of the magnetic fluxes in these
magnetograms{\footnote{Notice that the SoHO/MDI database we possess was
    constructed before the creation of the Level 1.8 MDI
    database. Therefore, underestimation of magnetic fluxes in
    sunspots and plage still exists in these data. Point taken,
    omitting the correction gives almost identical scaling constants $A$ and $\lambda$.}}. 
Substituting Equation (\ref{srel}) into Equation (\ref{Ec_lin}) for the
case of a single force-free flux tube we find 
\beq
E_c = A (\alpha d)^2 \Phi ^{2 \lambda}\;\;.
\label{Ec_single}
\eeq
The key assumption hereafter is that the scaling relation of Equation
(\ref{srel}) holds for {\it individual} flux tubes embedded into a
collection of discrete flux tubes with different $\alpha$-values. Indeed,
assuming that each magnetogram in Figure \ref{scat_plot} corresponds
to a flux tube with flux content $\Phi$ (the unsigned magnetic flux of the
magnetogram) and force-free parameter $\alpha$ (inferred by the linear
force-free approximation applied to the magnetogram), Equation
(\ref{Ec_single}) would provide the free magnetic energy of the flux
tube. This energy would be independent from the free energy of another
flux tube if mutual effects are ignored. Under these conditions,
we generalize Equation (\ref{Ec_single}) for a collection of $N$
flux tubes to provide the self term of the free magnetic energy of the
ensemble: 
\beq
E_{c_{self}} = A d^2 \sum _{l=1}^N \alpha _l^2 \Phi_l^{2 \lambda}\;\;.
\label{Ec_self}
\eeq
Equation (\ref{Ec_self}) can replace the respective
Equation (\ref{self1}) for $E_{c_{self}}$, hence accounting for the unknown 
$T_l^{close}$ in each flux tube $l$.

Regarding the self term of the relative magnetic helicity (Equation
(\ref{self1})), the linear force-free approximation of Paper I gives 
\beq
H_m = 8 \pi \mcal{F}_{lin}d^2 \alpha E_p\;\;.
\label{Hm_lin}
\eeq
Applying the scaling relation of Equation (\ref{srel}) we find 
\beq
H_m = 8 \pi d^2 \alpha A \Phi^{2 \lambda}\;\;,
\label{Hm_single}
\eeq
which is, again, independent for each flux tube if mutual terms are
neglected. Assuming that this scaling holds for individual flux tubes, we generalize 
Equation (\ref{Hm_single}) for a collection of $N$ flux tubes to obtain the self-helicity of the ensemble: 
\beq
H_{m_{self}} = 8 \pi d^2 A \sum _{l=1}^N \alpha _l \Phi_l ^{2 \lambda}\;\;.
\label{Hm_self}
\eeq
This can also replace the respective expression for $H_{m_{self}}$ in Equation
(\ref{self1}), thus accounting for the unknown $T_l$ in each flux tube $l$. 
\subsubsection{Mutual terms}
\label{S-method2b}
By construction, the linear analysis of Paper I cannot be used to
calculate any mutual energy or helicity terms. For these terms we will
implement the analysis of \citet{demoulin_etal_06}. 

Combining Equations (\ref{L_ij}) and (\ref{Ec_gen}), the mutual term of the magnetic free energy in our set of discrete, slender flux tubes is given by 
\beq
E_{c_{mut}}= {{1} \over {8 \pi}} \sum _{l=1}^N \sum _{m=1, l \ne m}^N 
          \alpha _l (\mcal{L}_{lm}^{close} + \mcal{L}_{lm}^{arch}) \Phi_l \Phi_m\;\;.
\label{Ec_mut}
\eeq
As a first step to calculate $E_{c_{mut}}$ we acknowledge that each term contributed by the interaction of a pair $(l,m)$ of flux tubes provides a positive increment $\Delta E_{c_{mut}}$ for $E_{c_{mut}}$. This increment is given by 
\beq
\Delta E_{c_{mut}}= {{1} \over {8 \pi}} (\alpha _l + \alpha _m) 
                  (\mcal{L}_{lm}^{close} + \mcal{L}_{lm}^{arch}) \Phi_l \Phi_m\;\;,
                  \;\;\;l, m \equiv \{1,...N \}\;\;.
\label{DEc_mut1}
\eeq
To provide Equation (\ref{DEc_mut1}) we have also taken into account that $\mcal{L}_{lm} = \mcal{L}_{ml}$, and this is the case for both $\mcal{L}_{lm}^{close}$ and $\mcal{L}_{lm}^{arch}$ and for every $(l,m)$. This simply means that the $N \times N$ helicity matrix is symmetric. Further, we know from theory that 
$\mcal{L}_{lm}^{close} = 0, \pm 1, \pm 2,...$ and 
$|\mcal{L}_{lm}^{arch}| < 1$ (Appendix A). In our pursuit for a minimum magnetic free energy, therefore, we assume that the interacting flux tubes do not wind around each other, so  $\mcal{L}_{lm}^{close} = 0$ for every $(l,m)$. This leads to a free-energy increment per interaction 
\beq
\Delta E_{c_{mut}}= {{1} \over {8 \pi}} (\alpha _l + \alpha _m) 
                  \mcal{L}_{lm}^{arch} \Phi_l \Phi_m\;\;,
                  \;\;\;l, m \equiv \{1,...N \}\;\;.
\label{DEc_mut2}
\eeq
According to \citet{demoulin_etal_06} and Appendix A there are only two possible values of 
$\mcal{L}_{lm}^{arch}$ in every possible case of interaction for a pair $(l,m)$ of flux tubes. These values depend on whether flux tube $l$ is "above'' flux tube $m$ ($\mcal{L}_{lm}^{arch} = \mcal{L}_{\hat{l}m}^{arch}$) or vice-versa 
($\mcal{L}_{lm}^{arch} = \mcal{L}_{l \hat{m}}^{arch}$). The possible cases of flux-tube interaction (see Figure \ref{mtx_fig}) are (i) intersecting footpoint segments of the interacting flux tubes, (ii) non-intersecting footpoint segments (there the two values coincide, i.e., 
$\mcal{L}_{\hat{l}m}^{arch} = \mcal{L}_{l \hat{m}}^{arch}$), and (iii) ``matching'' footpoint segments. A discussion of ``matching'' footpoints, where we assume that these footpoints are located within unresolved distances to raise the apparent physical inconsistency, is provided in Appendix A.

In summary, we have the following possibilities for $\Delta E_{c_{mut}}$: 
\bit
\item[$\bullet$] $\Delta E_{c_{mut}} >0$ for {\it one} possible $\mcal{L}_{lm}^{arch}$-value and $\Delta E_{c_{mut}} <0$ for the other. This is {\it always} true for intersecting footpoint segments and segments with ``matching'' footpoints. In this case we naturally select the $\mcal{L}_{lm}^{arch}$-value for which 
$\Delta E_{c_{mut}} >0$.
\item[$\bullet$] $\Delta E_{c_{mut}} >0$ for both $\mcal{L}_{lm}^{arch}$-values. This can only be found in case of non-intersecting footpoint segments, where 
 $\mcal{L}_{\hat{l}m}^{arch} = \mcal{L}_{l\hat{m}}^{arch}$. In this case, therefore, we naturally select this unique $\mcal{L}_{lm}^{arch}$-value.
\item[$\bullet$] $\Delta E_{c_{mut}} <0$ for both $\mcal{L}_{lm}^{arch}$-values. This also happens exclusively in cases of non-intersecting flux-tube footpoint segments, and is because the connectivity-matrix calculation of Section \ref{S-method1} relies only on $B_z$ and is hence independent of $\alpha$-values. One may envision an improved connectivity-matrix calculation in which the minimization functional of Equation (\ref{Mfunc}) is modified to include $\alpha$-values - this will aim toward connecting closely seated,  opposite-polarity partitions with like-sign force-free parameters. At the present stage, however, we cannot physically accommodate negative free-energy increments, so we enforce
 $\mcal{L}_{lm}^{arch}=0$ in these cases. Point taken, from practical experience this assumption does not incur a large change in the free energy $E_{c_{mut}}$ as this energy term is dominantly influenced by flux tube pairs with intersecting or "matching'' footpoint segments. 
\eit

Therefore, by means of Equation (\ref{DEc_mut2}) we both determine a unique 
$\mcal{L}_{lm}^{arch}$-value for all off-diagonal helicity-matrix elements and ensure a symmetric helicity matrix. Hence, our pursuit for a minimum free magnetic energy has also led to a corresponding value for the relative magnetic helicity. 

Can we reach an even smaller, but still positive, value for the mutual term of the free magnetic energy $E_{c_{mut}}$? The answer is yes, if we relax the assumption $\mcal{L}_{lm}^{close}=0$. In case $|\mcal{L}_{lm}^{arch}| >0.5$, for example, applying 
$|\mcal{L}_{lm}^{close}|=1$ such that $\mcal{L}_{lm}^{arch} \mcal{L}_{lm}^{close} <0$, one may find smaller, and in some cases positive, increments 
$\Delta E_{c_{mut}}$. Moreover, even for cases where $\Delta E_{c_{mut}} <0$ for both 
$\mcal{L}_{lm}^{arch}$-values and $\mcal{L}_{lm}^{close}=0$, we can always achieve 
$\Delta E_{c_{mut}} >0$ if we set $\mcal{L}_{lm}^{close}=1$ or $\mcal{L}_{lm}^{close}=-1$. All these possibilities are mathematically feasible. However, they lead to rather ``exotic'' physical situations of flux tubes winding around each other without necessarily intersecting footpoint segments. More importantly, they give rise to potentially uncontrollably high values of the mutual-helicity magnitude $|H_{m_{mut}}|$ or to helicity senses (chiralities) that run counter to the expected ones from observations of the active-region corona. We therefore follow the most conservative approach of keeping $\mcal{L}_{lm}^{close}=0$ and setting $\mcal{L}_{lm}^{arch}=0$ in case it only yields a negative $\Delta E_{c_{mut}}$. In essence, our approach suggests 
that a realistic state for a non-potential, force-free   
magnetic configuration is achieved when the free magnetic energy is
assumed minimum and the relative magnetic helicity is allowed to evolve in
this respect, considering only arched and not braided flux tubes. 
Whether solar magnetic fields are in a minimum free-energy
state is, of course, an open question. In Section \ref{S-val} we show, however, that validating our approach with existing, generally accepted
energy and helicity formulas leads to a fairly good agreement. 

Summarizing, given a collection of $N$ discrete, slender, force-free 
flux tubes with flux contents $\Phi _l$ and force-free parameters $\alpha _l$, where 
$l \equiv \{ 1,2,...N \}$, we write 
\ben
\item[(i)] the total magnetic energy as $E_t = E_p + E_c$, where $E_p$
  is the potential magnetic energy, and $E_c$ is the free magnetic
  energy. The latter term is given by 
\beq
E_c = A d^2 \sum _{l=1}^N \alpha _l^2 \Phi_l^{2 \lambda} + 
      {{1} \over {8 \pi}} \sum _{l=1}^N \sum _{m=1, l \ne m}^N 
           \alpha _l \mcal{L}_{lm}^{arch} \Phi_l \Phi_m\;\;.       
\label{Ec_fin}
\eeq
\item[(ii)] The relative magnetic helicity $H_m$ is given by 
\beq
H_m = 8 \pi d^2 A \sum _{l=1}^N \alpha _l \Phi_l ^{2 \lambda} + 
      \sum _{l=1}^N \sum _{m=1,l \ne m}^N \mcal{L}_{lm}^{arch} \Phi_l \Phi_m\;\;.
\label{Hm_fin}
\eeq
\een
The scaling constants $A$, $\lambda$ are given by the least-squares 
best fit of Equation (\ref{srel}), the mutual-helicity terms $\mcal{L}_{lm}^{arch}$ are explained and calculated in Appendix A, and $d$ is the size element. The force-free parameters $\alpha_l$ and flux contents $\Phi _l$ of the flux tubes are inferred
as described in Section \ref{S-method1}.

It should be mentioned here that the methodology we describe is more general than Equations (\ref{Ec_fin}), (\ref{Hm_fin}) in the sense that a nonzero Gauss linking number can always be accommodated in different physical settings. Moreover, our method can work for {\it any} given connectivity matrix. In this application, we physically favor both the connectivity matrix calculation of Section \ref{S-method1} and $\mcal{L}_{lm}^{close}=0$, thus reaching Equations (\ref{Ec_fin}), (\ref{Hm_fin}). 

A detailed error propagation analysis leading to the uncertainties 
$\delta E_c$ and $\delta H_m$ for $E_c$ and $H_m$, respectively, is
provided in Appendix B. 

In addition, one may infer the lowest possible free magnetic energy $E_{c_{WT}}$ that corresponds to a given amount of relative helicity for the NLFF field. This is simply the LFF free magnetic energy corresponding to this helicity, per the Woltjer-Taylor theorem \citep{woltjer_58, taylor_74, taylor_86}. To calculate $E_{c_{WT}}$ we use the NLFF relative helicity $H_m$ inferred by Equation (\ref{Hm_fin}) and we calculate an {\it effective} constant $\alpha$-value from Equation (\ref{Hm_lin}), i.e.
\beq
\alpha = {{H_m} \over {8 \pi d^2 \mcal{F}_{lin} E_p}}\;\;.
\label{al_wt}
\eeq
Substituting $\alpha$ from Equation (\ref{al_wt}) into Equation (\ref{Ec_lin}) for the LFF free energy, then, we obtain  
\beq
E_{c_{WT}} = {{H_m^2} \over {(8 \pi d)^2 \mcal{F}_{lin} E_p}}\;,\;\;\;or\;\;\;
E_{c_{WT}} = {{H_m^2} \over {(8 \pi d)^2 A \Phi ^{2 \lambda}}}\;\;,
\label{Ec_wt}
\eeq
due to Equation (\ref{srel}). 
The Woltjer-Taylor free magnetic energy $E_{c_{WT}}$ of Equation (\ref{Ec_wt}) will be used as sanity check in the following; all calculated free magnetic energies $E_c$ must be larger than this limiting value.  
\subsection{A special case: potential-field configurations}
\label{S-special}
Our choice to calculate the magnetic connectivity matrix via simulated annealing (Section \ref{S-method1}) implicitly assumes that the studied vector magnetograms are non-potential, that is, they include significant electric currents $I$ and force-free parameters $\alpha$ for {\it at least} one of the major partitions identified. Although the non-potentiality of solar active regions is a long-known fact \citep[e.g.,][]{zirin_wang_93, leka_etal_96}, our methodology includes a physical inconsistency in case a potential-field configuration, observed or modeled, is subjected to the analysis: the simulated annealing method, favoring strong magnetic polarity inversion lines, provides a connectivity that is generally  incompatible with that of the potential field. The question to ask, then, is how to determine whether a vector magnetogram is basically potential. 

One might argue that mutual-helicity terms $\mcal{L}_{lm}^{arch} \Phi_l \Phi_m$ algebraically cancel to zero in case of potential fields, along with individual self-helicity factors $T_l$ that tend to zero -- otherwise, the relative-helicity expression of \citet{demoulin_etal_06} (Equation (\ref{Hm_gen})) and our final formula 
(Equation (\ref{Hm_fin})) are not valid. In practice, however, an observed active-region magnetogram is not flux-balanced. This will inhibit an algebraic cancellation of mutual-helicity terms that may sum up to significant nonzero helicity values. 
At the same time, our expression for the free magnetic energy 
(Equation (\ref{Ec_fin})) will give values close to zero because $\alpha \simeq 0$ for all partitions, and hence for all possible flux-tube connections. Hence, the physical inconsistency mentioned above leads to another inconsistency, namely to situations of near-zero magnetic free energy and strongly nonzero relative helicity.  

There are several methods to determine the degree of non-potentiality in observed active regions. We are currently in the process of identifying the most viable and practical of them, that will be reported in a future publication. One of these methods is to compare the observed horizontal-field components with those of the potential field obeying to the observed vertical-field distribution. Such a comparison may rely on scaling indices, correlation coefficients, and/or standard deviations between the observed and modeled components. Another method is to provide a flux-weighted mean $\alpha$-value $\bar{\alpha}$ and a respective uncertainty $\delta \bar{\alpha}$ from the $\alpha$-values of all partitions. In case 
$|\bar{\alpha}| \le n_{\sigma} \delta \bar{\alpha}$, where $n_{\sigma}$ is a given significance level, the active region may be considered potential. Further, a flux-weighted mean magnetic-shear angle in the region is certainly another index of non-potentiality. In case of a nearly potential active-region magnetogram, rather than performing simulated annealing, one might perform a potential-field extrapolation and infer the connectivity matrix by line-tracing of the resulting magnetic field lines. Alternatively, one might stop the calculation at this point and set $E_c=0$ and $H_m=0$ for the active region of interest, thus saving computing time and avoiding large uncertainties owning to the magnetic-flux imbalance.   
\section{Method validation}
\label{S-val}
\subsection{Benchmarking: volume formulas for magnetic energy
and helicity}
\label{S-val1}
Well-known formulas for magnetic energy and helicity can be used 
in case one knows the three-dimensional magnetic configuration above a
two-dimensional planar boundary, where the magnetic field vector is
fully known. For a three-dimensional flux-balanced magnetic structure the energy budgets are 
\beq
E_t = {{1} \over {8 \pi}} \int _{\mcal{V}} B^2 d \mcal{V}\;,\;\;\;
E_p = {{1} \over {8 \pi}} \int _{\mcal{V}} B_p^2 d\mcal{V}\;,and\;\;\;
E_c = E_t - E_p\;\;,
\label{Et_vol}
\eeq
where $B=|\mbf{B}|$ is the magnetic field strength in the calculation
volume $\mcal{V}$ and $B_p=|\mbf{B_p}|$ is the respective (also fully
known) field strength of the potential field in $\mcal{V}$, where $\mbf{B}$ and $\mbf{B_p}$ share the same normal component on the lower boundary. 

For the relative magnetic helicity $H_m$ in $\mcal{V}$,  
\citet{berger_field_84} and \citet{finn_antonsen_85} derived two 
equivalent analytical forms valid for NLFF fields, namely 
\beq
H_m = \int _{\mcal{V}} (\mbf{A} \pm \mbf{A_p}) \cdot 
                      (\mbf{B} \mp \mbf{B_p}) d \mcal{V}\;\;,
\label{Hm_vol}
\eeq
where $\mbf{A_p}$ and $\mbf{A}$ are the vector potentials for
$\mbf{B_p}$ and $\mbf{B}$, respectively. Although $\mbf{B_p}$ and
$\mbf{B}$ may be exactly known, proper knowledge of $\mbf{A_p}$, and 
especially $\mbf{A}$, is a much more demanding task. Point taken,
the substantial value of Equation (\ref{Hm_vol}) is that one may use
two gauge-dependent, non-unique expressions for $\mbf{A_p}$ and
$\mbf{A}$ to obtain a gauge-invariant, unique expression for $H_m$
(see also \citealt{berger_99}). This paved the way for
implementations such as the vector-potential expressions of 
\citet{devore_00} and \citet{longcope_malanushenko_08}, 
among others. These
expressions, however, are valid for the semi-infinite space
(half-space) above the lower-boundary plane. Recently,
\citet{valori_etal_11} showed that if the vector potentials $\mbf{A_p}$
and $\mbf{A}$ are corrected for the fact that the calculation volume
$\mcal{V}$, wherein $\mbf{B_p}$ and $\mbf{B}$ are known, is {\it finite},
then the resulting $H_m$-values from Equation (\ref{Hm_vol}) may be
significantly different in amplitude and sometimes even in sign. In
particular, these authors extended the analysis of \citet{devore_00} and 
found that if $\mcal{V}$ extends between $(x_1, x_2)$, $(y_1, y_2)$,
$(z_1, z_2)$ in a cartesian coordinate system, then
\beq
\mbf{A_p} = \mbf{c} - \nabla \times (\hat{z} \int _{z_1}^z \phi dz')\;\;,
\label{Ap_vol}
\eeq
where $\phi$ is the scalar potential generating $\mbf{B_p}$
($\mbf{B_p} = - \nabla \phi$) and $\mbf{c}=(c_x,c_y)$, with 
\begin{eqnarray}
\begin{array}{lll}
c_x & = & -(1/2) \int _{y_1}^y B_{p_z} (x,y',z=z_1) dy'\\
c_y & = &  (1/2) \int _{x_1}^x B_{p_z} (x',y,z=z_1) dx'\;\;,\\
\end{array}
\label{cxy}
\end{eqnarray}
and
\beq
\mbf{A} = \mbf{A_p}(x,y,z=z_1) - \hat{z} \times \int _{z_1}^z \mbf{B} dz'\;\;,
\label{A_vol}
\eeq
by choosing a gauge such that 
$\hat{z} \cdot \mbf{A_p} = \hat{z} \cdot \mbf{A} =0$ everywhere in
$\mcal{V}$. 

To test our NLFF expressions 
we will apply the derivations of \citet{valori_etal_11} for $\mbf{A_p}$
and $\mbf{A}$ to Equation (\ref{Hm_vol}) for relative helicity 
in cases where we have
performed NLFF field extrapolations on photospheric vector
magnetograms. From these extrapolations and their potential-field
counterparts we will also calculate the magnetic energy budgets of
Equations (\ref{Et_vol}). To be perfectly consistent with the purposes
of this validation test, the connectivity matrix $\mathscr{F}^{con}_{ij}$ will 
not be inferred by simulated annealing (in this particular case only), but by tracing the NLFF-field-extrapolated lines. 
\subsection{Validation results}
\label{S-val2}
Validating our NLFF energy and helicity calculation method is a nontrivial exercise. At first glance, one might rely on analytical NLFF field models 
\citep[e.g.,][]{low_lou_90, regnier_etal_05, regnier_11}.
However, there are two shortcomings in this approach: first, in some of these models the vector potential is unknown, so one still needs to calculate gauge-dependent values of it to infer helicity. 
Second, and most importantly, the lower-boundary configuration is very simple in these models resulting in a very limited magnetic connectivity matrix. As we intend to emphasize the practical aspect of the methodology presented here, we choose to validate the method on {\it real} vector magnetograms, using generally accepted energy and helicity formulas applicable to the coronal volume above the lower-boundary magnetogram (Section \ref{S-val1}). This volume is determined by a NLFF field extrapolation. We use the cartesian optimization method developed by 
\citet{wiegelmann_04}, where the divergence of the magnetic field vector and the Lorentz force are simultaneously minimized for the configuration to converge to a NLFF state. No preprocessing of the boundary vector magnetogram 
\citep{wiegelmann_etal_06} was attempted in this case. 

Browsing through a sizable collection of vector magnetograms \citep{tziotziou_etal_12} we carefully selected 19 of them in which (i) the fractional magnetic flux imbalance is relatively low, thus allowing the majority of the unsigned flux to participate in the connectivity matrix, and (ii) the NLFF field extrapolation has worked well, with acceptable minimizations of the divergence of the field vector and the Lorentz force, and with a convergent solution exhibiting a total magnetic energy larger than the potential-field energy. To ensure that the extrapolated three-dimensional field solution is valid we consider the dimensionless parameter 
\beq
\mathcal{D} = { {|\nabla \cdot B|} \over {\sqrt{3} 
 [ (\partial B_x/\partial x)^2+ (\partial B_y/\partial y)^2 + 
      (\partial B_z/\partial z)^2  ]^{1/2}} }\;\;.
\label{dcheck}
\eeq
The flux-weighted mean of $\mathcal{D}$ for all used magnetograms is 
$\lesssim 0.1$ indicating roughly divergence-free, and hence valid, field solutions. The data were acquired by the IVM, with a binned pixel size of $2.2\arcsec$, and the Spectropolarimeter (SP; \citealt{lites_etal_08}) of Hinode's Solar Optical Telescope (SOT; \citealt{tsuneta_etal_08}). SOT/SP is a spectrograph observing in two magnetically sensitive photospheric spectral lines; Fe 
{\small I} at 6301.5 \AA$\;$ and 6302.5 \AA, respectively, with a spectral sampling of $21.6$ m\AA$\;$. Full instrument resolution ($\sim 0.32\arcsec$) corresponds to a pixel size of $\sim 0.158\arcsec$. The data used here, however, have been acquired in fast scanning mode, corresponding to $\sim 0.308\arcsec$ per pixel.  We have resolved the azimuthal $180^o$ ambiguity in these data using the Non-Potential Field Calculation (NPFC) method of \citet{georgoulis_05} - see also \citet{georgoulis_11a}. Then, to further expedite the 
extrapolations, we have spatially binned the data to $\sim 1.25\arcsec$ per pixel, indicating a spatial resolution of $\sim 2.5\arcsec$. \citet{lites_etal_08} reported line-of-sight and transverse-field uncertainties equal to 2.4 $Mx\;cm^{-2}$ and 41 $Mx\;cm^{-2}$, respectively, for the quiet Sun. In the following calculations we use uncertainties of $5\;G$ and $50\;G$, respectively, for SOT/SP, and $50\;G$ and $100\;G$, respectively, for IVM data. The same uncertainties apply for the calculations performed in the following Sections. 

Each vector magnetogram is then subjected to both volume-integral energy and helicity calculations (Section \ref{S-val2}) and the NLFF surface calculations of this work. A crucial point here is that, for a direct comparison between the resulting energy and helicity budgets, one must use the connectivity matrix inferred by the NLFF field extrapolations. This is what we have done for this test only, inferring the various connectivity matrices by line-tracing the respective extrapolation results. The results of the comparison for the relative magnetic helicity and the free magnetic energy are given in Figures \ref{val_plot}a and \ref{val_plot}b, respectively. Error bars correspond to uncertainties calculated by our NLFF field method while the red lines denote equality between the compared budgets. 

In terms of the relative magnetic helicity magnitude (Figure \ref{val_plot}a) we notice a fairly good agreement between the volume-calculated and the surface-calculated values -- 
most points are within uncertainties from the equality line. The inferred helicity sense for extrapolations and our surface calculations agree in all 19 cases. For relatively small relative helicities ($\lesssim 10^{42}$ $Mx^2$), our method appears to overestimate the relative helicity, albeit generally within applicable error bars. The linear (Pearson) and rank order (Spearman) correlation coefficients are significant, ranging between 0.63 and 0.74. 

The results of the comparison in magnetic free energy (Figure \ref{val_plot}b) are similar to that of the relative helicity magnitudes. For small free energies 
($\lesssim 10^{31}$ erg), our surface calculation seems to overestimate, within error bars, the respective values while the opposite, beyond error bars, occurs for free energies 
$> 10^{32} erg$. The Pearson and Spearman correlation coefficients are also significant, higher than in the case of relative helicity, and ranging between 0.74 and 0.75. We therefore conclude that the approximation of the free magnetic energy with the expression of Equation (\ref{Ec_gen}) is a reasonable choice, despite shortcomings, given the circumstances and the incomplete information. 

It is worth noting at this point that discrepancies between the volume and the surface calculation of magnetic free energy and relative helicity may be due to errors and ambiguities inherent to {\it both} our calculations, given the assumptions adopted, and the NLFF field extrapolations, including the calculation of the vector potentials that participate in the relative helicity formula of Equation (\ref{Hm_vol}). Given the numerical methods involved, the derived vector potentials $\mbf{A_p}$ and $\mbf{A}$  reproduce the magnetic field vectors $\mbf{B_p}$ and $\mbf{B}$, respectively, within non-negligible differences.    

In Figure \ref{val_plot}b we have also plotted the Woltjer-Taylor free magnetic energy $E_{c_{WT}}$ of Equation (\ref{Ec_wt}) (crosses). These energies are to be viewed as sanity checks since the NLFF-field free magnetic energy $E_c$ cannot be smaller than them. In only one case in Figure \ref{val_plot}b does $E_{c_{WT}}$ exceed $E_c$ (fourth point from left). This, however, is within the applicable error bar. We further notice that some $E_{c_{WT}}$-values are up to $\sim 2$ orders of magnitude smaller than $E_c$, hence being unrealistically low, while some are quite close to their respective $E_c$-values. This may suggest the plausibility of the Taylor relaxation in at least some active regions. Although this discussion exceeds the scope of this work, we note in passing that the validity of the Taylor relaxation in the solar corona, according to which a relatively isolated magnetic configuration may relax in a state of minimum free (LFF-field) magnetic energy for a given magnetic helicity budget, is still a subject of debate. Perhaps calculation methods such as the one proposed here can provide further clues to judge the validity of 
this hypothesis. 

In brief, we conclude that our surface-based NLFF field calculation method manages to provide {\it both} magnetic free energies {\it and} relative helicities in fairly good agreement with generally accepted, but computationally much more intensive and model-dependent, volume-calculation techniques. It is, therefore, a viable method that can be exploited further and in larger data sets of solar vector magnetograms. 
\section{Results: NLFF field energy and helicity calculations}
\label{S-res}
\subsection{NOAA ARs 8844 and 9165, in comparison with Paper I}
\label{S-res1}
For the purposes of comparison with Paper I we use here the same two vector magnetogram timeseries, namely those of NOAA ARs 8844 and 9165, both acquired during daily observing cycles of the IVM on 2000 January 25 and September 15, respectively. The IVM data acquisition and selection and the properties of each AR are described in detail in Section 5.1 of Paper I. In brief, NOAA AR 8844 was a small emerging flux region, visible in the solar disk between 2000 January 24 and 27. No significant eruptive activity originated from this AR. On the other hand, NOAA AR 9165 was a complex, persistent, and eruptive region hosting a number of eruptive M-class flares (see Figure 2 of Paper I). Both ARs were fairly well flux-balanced within the IVM field of view: flux imbalance was $\lesssim 5$\% for NOAA AR 8844 and $\lesssim 10$\% for NOAA AR 9165 (Figure 3 of Paper I). The mean unsigned flux was 
$\sim 5.1 \times 10^{21}$ Mx for NOAA AR 8844 and $\sim 17 \times 10^{21}$ Mx for NOAA AR 9165, so with a unsigned-flux ratio $\sim 3.4$. Perhaps more relevant in this case is the total flux $\sum \mathscr{F}^{con}_{ij}$ that participates in 
the connectivity matrices (Section \ref{S-method1}), where NLFF free energies and helicities stem from. The mean connected-flux values are $\sim 3.6 \times 10^{21}$ Mx and 
$\sim 12.8 \times 10^{21}$ Mx for NOAA ARs 8844 and 9165, respectively, so with a connected-flux ratio of $\sim 3.6$. We notice that most of the unsigned flux for both ARs participates in the connectivity matrices. This is because both ARs are nearly flux-balanced. The little remaining flux that does not participate is either too disperse to be included in the flux partitioning or is judged to be connected with opposite-polarity flux concentrations seated beyond the field of view. 

Figure \ref{8844_comp} provides the timeseries of the calculated NLFF relative helicity $H_m$ (Figure \ref{8844_comp}a) and free energy $E_c$ (Figure \ref{8844_comp}b) for NOAA AR 8844. For an immediate comparison we have also plotted the respective LFF values from Paper I (gray curves and shades)
while the NLFF field calculations of this work are shown with blue curves and shades. Assuming that within the observing interval of $\sim 2$ hours there was no significant change in $H_m$ and $E_c$ we define mean values $\bar{H}_m$ and $\bar{E}_c$ for $H_m$ and $E_c$, respectively, accompanied by the respective standard deviations. 
We find $\bar{H}_{m_{NLFF}} = (1.18 \pm 0.45) \times 10^{42}\;Mx^2$ and 
$\bar{E}_{c_{NLFF}} = (1.52 \pm 0.42) \times 10^{31}\;erg$ in the NLFF field  approximation. 
Mean values are shown by the blue and gray straight lines for the NLFF and LFF calculation, respectively, while the respective standard deviations are shown by the blue- and gray-shaded areas. We notice that (i) the LFF and the NLFF approaches give values of $\bar{H}_m$ and $\bar{E}_c$ that are fairly close to each other (generally within uncertainties) at least for this small AR, and (ii) the NLFF field approximation gives an overall smoother evolution with smaller standard deviations. A summary of the values and uncertainties for 
the NLFF $\bar{H}_m$ and $\bar{E}_c$ is provided in Table \ref{Tb1}.

Figure \ref{9165_comp} provides the timeseries of the calculated NLFF relative helicity $H_m$ (Figure \ref{9165_comp}a) and free energy $E_c$ (Figure \ref{9165_comp}b) for NOAA AR 9165. In this case the mean values $\bar{H}_m$ and $\bar{E}_c$ are higher than in NOAA AR 8844  
($\bar{H}_{m_{NLFF}} = (-7.3 \pm 1.4) \times 10^{42}\;Mx^2$ and 
$\bar{E}_{c_{NLFF}} = (5.3 \pm 1.4) \times 10^{31}\;erg$). The differences between the LFF- and the NLFF-field approximations in NOAA AR 9165 are also larger: in, general, the LFF-field approximation tends to overestimate {\it both} the magnitude of the relative magnetic helicity and the free magnetic energy. The overestimation caused by adopting the LFF-field approximation is $\sim 1.8$ for $\bar{H}_m$ and $\sim 2.5$ for $\bar{E}_c$. This is understandable and expected as the LFF-field approximation assigns a fixed force-free parameter $\alpha$ with a value determined by the strongest (most flux-massive) non-potential field configurations in the AR (see the analysis of deriving a single maximum-likelihood $\alpha$-value in Paper I). In addition, the effect of obtaining smaller uncertainties for $\bar{H}_m$ and $\bar{E}_c$ in the NLFF-field approximation is more evident in NOAA AR 9165. A summary of the mean values and respective uncertainties for NOAA AR 9165 is also provided in Table \ref{Tb1}.

Comparing the mean values $\bar{H}_m$ and $\bar{E}_c$ for the two studied NOAA ARs 8844 and 9165 we find a ratio of $6.2 \pm 2.9$ for $\bar{H}_m$ and $\sim 3.5 \pm 1.3$ for 
$\bar{E}_c$ between the flaring and the non-flaring regions. These ratios are roughly similar to the unsigned- and connected-flux ratios ($\sim 3.4$ and $\sim 3.6$, respectively) but smaller than those reported in Paper I for the LFF-field  calculations of $\bar{H}_m$ and $\bar{E}_c$. In that work, both ratios were $\sim 9$. We conclude that the overestimation of the magnetic free energy and relative magnetic helicity in the LFF field approximation is higher for larger, more complex active regions. Point taken, the flaring NOAA AR 9165 still shows much larger free-energy and relative-helicity budgets compared to the flare-quiet NOAA AR 8844. This quantitative distinction between eruptive and non-eruptive active regions is studied by \citet{tziotziou_etal_12}. 

Concluding our calculations on NOAA ARs 8844 and 9165, we comment on the
contributions of the self and mutual terms to the free energy and relative helicity budgets (Equations (\ref{Ec_fin}), (\ref{Hm_fin}), respectively) in the two studied ARs. We find that mutual terms overwhelmingly dominate these budgets: on average, for NOAA AR 8844 self terms contribute $(0.1  \pm 0.07)$\% of the free magnetic energy and $(0.1 \pm 0.2)$\% of the relative magnetic helicity. The respective percentages for NOAA AR 9165 are $(0.5 \pm 0.2)$\% and $(0.4 \pm 0.2)$\%. 
These findings are in qualitative agreement with previous works 
\citep{regnier_etal_05, regnier_canfield_06} and suggest that twist and writhe (contributing to self helicity) are {\it numerically} far less important than the mutual helicity caused by the interaction between different  flux tubes. 
\subsection{NOAA AR 10930}
\label{S-res2}
An appealing aspect of our analysis is that it can be applied to 
{\it long-term timeseries} of active-region vector magnetograms in order to reveal the temporal variation of magnetic energy and helicity budgets in the studied regions. The input vector magnetograms for this purpose should ideally exhibit constant quality and a fixed, high cadence. This is a big challenge for aging  ground-based magnetographs such as the IVM. The Vector Spectromagnetograph (VSM; \citealt{henney_etal_09}) of the SOLIS facility \citep{keller_etal_03} has achieved important  advances on the quality front but, by design, it does not exhibit a cadence higher than a few hours. The space-based 
SOT/SP onboard Hinode exhibits unprecedented spatial resolution and a 
constant data quality due to the lack of atmospheric interference but, again, by design it only allows a cadence of a few hours. A lasting solution will have been achieved when the Helioseismic and Magnetic Imager (HMI; \citealt{scherrer_etal_11}) onboard the Solar Dynamics Observatory (SDO) releases constant-quality vector magnetograms of solar active regions with a fixed cadence of 12 minutes (see \citet{sun_etal_12} for an example). At this time, however, for the purpose of showing the magnetic energy and helicity variations in an active region over a period of days, we present results obtained by processing a timeseries of SOT/SP vector magnetograms of NOAA AR 10930. 

NOAA AR 10930 appeared in the earthward solar hemisphere in 2006 December. It was an intensely flaring (and eruptive) region hosting $\sim 50$ C-class, 5 M-class and 4 X-class flares before rotating beyond the western solar limb. The AR has been studied in extreme detail by dozens of works; in some of them estimates of the magnetic free energy and the relative magnetic helicity in the region have been published. For example, \citet{ravindra_etal_11} reported that, by 2006 December 13, when the AR hosted a X3.4 flare, more than 
$-6 \times 10^{43}$ $Mx^2$ of relative helicity had been injected in the AR. A slightly more conservative estimate was published by \citet{park_etal_10}, with a relative helicity $\sim -4.3 \times 10^{43}$ $Mx^2$ before the flare. As to the free magnetic energy of the AR, \citet{he_etal_11} estimated it within  
$(1.25 - 1.4) \times 10^{33}$ erg, while \citet{guo_etal_08} reported a value of 
$\sim 2 \times 10^{32}$ erg.

We have selected 30 vector magnetograms of the region acquired by Hinode's SOT/SP between 2006 December 8 and December 14. The heliographic vertical magnetic field component of six of them is shown in Figure \ref{10930_im} with Figures \ref{10930_im}a and \ref{10930_im}f corresponding to the first and last magnetogram of the series, respectively. The magnetograms were first disambiguated using the NPFC method and then were spatially binned to $\sim 0.62\arcsec$ per pixel, or to a spatial resolution of $\sim 1.23\arcsec$. 

The evolution of the relative magnetic helicity in NOAA AR 10930 is shown in Figure \ref{10930_calc}a. A distinct feature of the pre-eruption evolution in the AR is that, before 12/10, the relative-helicity budget is rather low, of the order $(-5 \pm 3) \times 10^{42}$ $Mx^2$. Over the next two days (12/11 - 12/12), however, the helicity budget increases drastically to reach 
$\sim -1.1 \times 10^{43}$ $Mx^2$. This peak coincides with a cluster of C-class flares (see the respective GOES 1-8 \AA$\;$X-ray flux in Figure \ref{10930_calc}d) that appear to be eruptive, as can be judged by the repetitive Type-II bursts recorded in the frequency-time radio spectra of the WAVES instrument onboard the WIND mission \citep{bougeret_etal_95}. Type II activity implies shock-fronted CME occurrences \citep[e.g.,][]{nelson_melrose_85}. Although neither GOES nor WIND/WAVES have spatial resolution, there is little doubt that the observed eruptions originate from NOAA AR 10930, as it was the only AR present in the visible solar disk at the time. The CMEs and their locations are also confirmed by the SoHO/LASCO CME database \citep{yashiro_etal_04}. 

Around the start of 12/12, perhaps due to eruptive activity, the helicity budget appears generally smaller, of the order $-(7 \pm 3) \times 10^{42}$ $Mx^2$. Hours before the X3.4 flare, early on 12/13, however, the relative helicity budget increases substantially to exceed $\sim -1.5 \times 10^{43}$ $Mx^2$. At this time of peak helicity the eruptive flare occurs, accompanied by a fast halo CME. Immediately after the eruption, late on 12/13, 
the AR appears to have lost 
$\sim -5 \times 10^{42}$ $Mx^2$ of helicity, perhaps in the CME. For a period of $\sim 12$ hours until the end of the observing interval the helicity budget appears to be of the order $\sim -1 \times 10^{43}$ $Mx^2$.

Around the time of the X-class flare we calculate a relative magnetic helicity budget that is a factor of $\sim 3$ lower than the estimate of \citet{park_etal_10} and a factor of $\sim 4$ lower than the estimate of 
\citet{ravindra_etal_11}. Our lower helicity estimate is consistent with (i) the fact that we calculate helicity from closed-field connections only, thus using only a fraction of the unsigned flux (a small fraction in this case, as the AR shows significant flux imbalance -- Figure \ref{10930_calc}c) and (ii) our methodology, that minimizes the free magnetic energy first, and then keeps a  consistent helicity magnitude. In addition, our study is qualitatively consistent with the assessment of \citet{park_etal_10} that helicity magnitude abruptly increases on 12/10 and thereafter. A detailed discussion of the process exceeds the scope of this work that mainly focuses on the proposed calculation methodology. 

The evolution of the magnetic energy budgets in NOAA AR 10930 is shown in Figure 
\ref{10930_calc}b. Here we notice that while the potential energy is linearly increasing over the observing interval, in agreement with the increase of the unsigned flux (Figure \ref{10930_calc}c), the free magnetic energy increases with a faster rate on 12/10 and thereafter, when the relative helicity magnitude increases. The peak helicity at the time of the multiple eruptive C-class flares coincides with a free energy of $\sim 2 \times 10^{32}$ erg. The free energy is kept at approximately these levels, albeit showing a moderate increasing trend, until late on 12/12. Thereafter, it increases significantly to peak at $\sim 5 \times 10^{32}$ erg around the time of the X-class flare. After the flare, following the decrease of the helicity budget, the free energy decreases to $\sim 3 \times 10^{32}$ erg, to remain at these levels until the end of the observing interval. Given the small uncertainties, the decrease of $\sim 2 \times 10^{32}$ erg in the course of the eruption appears significant. 

Our free-energy estimates agree qualitatively with those of \citet{guo_etal_08} but they are significantly lower than those of \citet{he_etal_11}. This is in agreement with our effort to keep the free energy minimum, at the same time excluding from calculation all magnetic connections that close beyond the field of view. This being said, the postflare free energy decrease is quite consistent with \citeauthor{hudson_11}'s (\citeyear{hudson_11}) assessment that X-class flares typically dissipate $\sim 10^{32}$ erg of magnetic energy.

In summary, we find that a plausible physical interpretation of the dynamical evolution of the eruptive NOAA AR 10930 can rely on our calculation of the relative magnetic helicity and free magnetic energy in the region. Moreover, the estimated magnetic energy and relative helicity budgets are consistent with the lowest published estimates, as expected. Given that our calculations (i) stem from a {\it unique} solution for the magnetic connectivity matrix, (ii) do not depend on magnetic field extrapolations, and (iii) are relatively inexpensive computationally, we argue that the proposed method is both viable and practical. 
\section{Discussion and Conclusions}
\label{S-fin}
Motivated by the need to achieve a practical, realistic, and self-consistent assessment of the magnetic energy and relative magnetic helicity budgets in solar active regions we first tackled the problem in the simplified LFF field approximation (\citealt{georgoulis_labonte_07} -- Paper I). In this case the solution was unique and dependent on the single, fixed force-free parameter used. However, it is known that there are multiple ways to infer this single $\alpha$-value 
\citep[i.e.,][]{leka_skumanich_99, leka_99} and the one used in Paper I was but one of several methods. Different $\alpha$-values give different solutions for the magnetic energy and helicity budgets. Moreover, the LFF field approximation is generally unrealistic in active-region scales (see, however, \citet{moon_etal_02} for a different view). For this reason the analysis of Paper I was conceived as the first step toward a more realistic approach of performing magnetic energy and helicity calculations. This step, relying on a NLFF field approximation, is taken in this work. 

Multiple methods to calculate the magnetic energy and relative helicity budgets of solar active regions assuming NLFF magnetic fields are long present in the literature. Virtually all of them rely either on a photospheric velocity flow field or on a three-dimensional NLFF field of the active-region corona (for a review, see \citet{demoulin_07} and references therein). Both the photospheric flow field and the three-dimensional coronal NLFF field, however, are not uniquely defined. Therefore, the resulting NLFF-field energy and helicity budgets are, again, model-dependent. An apparently more robust method to calculate the 
{\it spinning} and {\it braiding} helicity in observed solar active regions was introduced by \citet{longcope_etal_07} and applied to NOAA AR 10930 by 
\citet{ravindra_etal_11}. The method also uses a velocity field but now this field is obtained by tracking the motions of photospheric flux partitions. These partitions are  inferred as described in Section \ref{S-method1} and Figure \ref{mtx_fig}. Feature tracking on individual partitions should be more stable than calculating the entire flow field, although limitations have been reported for this case, as well. 

Here we follow a different approach depending on neither photospheric flow fields nor the unknown coronal three-dimensional field. Our method depends on a lower-boundary (photospheric or chromospheric) magnetic connectivity matrix that can be inferred either by a NLFF field extrapolation or otherwise. Had we used an extrapolation, our results would also be model-dependent. Although our method is general enough to accommodate {\it any} connectivity matrix, we propose and use a 
{\it unique} connectivity-matrix solution for a given flux-partition map. This solution relies on a simulated-annealing algorithm designed to minimize the distances of connected opposite-polarity partitions, thus emphasizing strong polarity inversion lines. This connectivity methodology has been successful in distinguishing flaring from non-flaring active regions \citep{georgoulis_rust_07} and will be shown to be of further such importance in a much larger statistical sample (work in preparation). Besides the boundary connectivity, our method alleviates the need for flow fields and three-dimensional coronal field vectors by, first, minimizing the magnetic free energy of the active-region corona and, second, keeping the dominant mutual-helicity terms consistent with the global (within active-region scales) energy-minimization principle. Self terms of magnetic energy and helicity are calculated by generalizing the LFF field analysis of Paper I, while mutual terms are calculated by a practical implementation of the method introduced by \citet{demoulin_etal_06}. 

To validate the proposed method we use connectivity matrices derived from NLFF field extrapolations because comparison is then based on the NLFF model-dependent energies and helicities. For the validation we use real active-region magnetograms, thus avoiding synthetic NLFF fields with a simpler lower boundary and hence a smaller and cruder connectivity matrix. At the same time, the approximate validity of the volume-calculated energies and helicities is guaranteed by the use of well-known and accepted energy and helicity formulas. We find (Section \ref{S-val2}) that the results of known volume formulas that require a detailed three-dimensional field configuration are generally reproduced by our surface formulas that use only the connectivity matrices inferred from the extrapolations. This justifies the use of the free-magnetic-energy formula of 
Equation (\ref{Ec_fin}) as a lower limit of the true free energy, in spite of its weakness to fully describe space-filling, force-free fields (Section \ref{S-method2}).

Following validation, we apply our method to the same set of active regions with Paper I and compare the results (Section \ref{S-res1}). We find them to be consistent, in general, but with the LFF field approximations overestimating the free magnetic energy and relative magnetic helicity budgets in such a way that overestimation is higher for larger, more complex active regions. In the NLFF field approximation the ratios of the free magnetic energy and the relative magnetic helicity between the two ARs are roughly similar to the ratio of the unsigned flux participating in the magnetic connectivity matrices ($\bar{\Phi}_{MTX}$ -- see Table \ref{Tb1}), as opposed to $\sim \bar{\Phi}^2$, i.e., the square of the unsigned magnetic flux, obtained by the LFF field approximation of Paper I. In addition, we find that mutual-energy and helicity terms overwhelmingly dominate the respective budgets in both active regions with contributions in excess of 99.5\%. 

Our NLFF field approach is then applied to a timeseries of vector magnetograms of the eruptive NOAA AR 10930, observed over a period of $\sim 6$ days (Section \ref{S-res2}). The results corroborate the findings of \citet{ravindra_etal_11} that the relative helicity in the region increased abruptly within $\sim 24$ hours, resulting in significant left-handed helicity in the region. The peak helicity magnitude we find, however, is $\sim 4$ times smaller than that of \citet{ravindra_etal_11} and $\sim 3$ times smaller than that of \citet{park_etal_10}. We also find that (i) an initial helicity magnitude decrease is associated in time with a cluster of eruptive C-class flares, and (ii) a more abrupt helicity decrease of $\sim -5 \times 10^{42}$ $Mx^2$ relates closely in time with the eruptive X3.4 flare that climaxes the eruptive activity in the region over the observing interval. A similar decrease of $\sim 2 \times 10^{32}$ erg in the region's free magnetic energy was also calculated at that time. Both the relative helicity and the free energy remained at these lower-budget levels until the end of the observing interval. To firmly establish these findings, however, more vector magnetogram data are necessary. 

The practical energy and helicity calculations in observed solar active regions being the scope of this work, the ultimate objective, also posed in Paper I, is to acknowledge and outline the possible role of magnetic helicity in the triggering of solar eruptions. We take a first step in this direction in 
\citet{tziotziou_etal_12}, where the method introduced here is applied to 162 active-region vector magnetograms to yield the first {\it energy-helicity diagram} of solar active regions. This diagram demonstrates a monotonic correlation between the free magnetic energy and the relative magnetic helicity in active regions, at the same time showing a segregation between flaring/eruptive regions and non-eruptive ones. This finding reinforces the results of previous works in observed active regions \citep{nindos_andrews_04, labonte_etal_07, georgoulis_etal_09}
and in theory (\citealt{zhang_low_01, zhang_low_03} -- see also \citet{nindos_09} for a review) that eruptions leading to CMEs effectively transfer excess magnetic helicity from the Sun outward to the heliosphere. One is tempted to assert, therefore, that if CMEs are means to relieve the Sun from its excess helicity, then helicity itself may play a key role in solar eruptions. Other than the helical kink instability, known to lead to eruptions in {\it some} observed filament destabilizations \citep{rust_labonte_05}, however, and the helicity-annihilation mechanisms of \citet{kusano_etal_03} that remains to be proved, the alleged role of helicity is unknown. Only very recently, \citet{kliem_etal_12} demonstrated that weakly kink-unstable magnetic configurations can represent observed solar-eruption features. Moreover, \citet{raouafi_etal_10} presented evidence that several X-ray jets observed by Hinode's X-Ray Telescope (XRT; \citealt{golub_etal_07}) are preceded my micro-sigmoids, thus elucidating the possible role of the helical kink instability even in small-scale eruptive activity. Finally, \citet{patsourakos_vourlidas_09} stereoscopically 
observed a small transient sigmoid that erupted giving rise to an observed EIT wave. In view of the above and other results, a detailed investigation of the role of helicity in the pre-eruption configuration of eruption-prolific solar active regions is well justified. Such an investigation was sketched by \citet{georgoulis_11b} and will be the subject of a study currently in preparation.
\acknowledgements
We thank T. Wiegelmann for the kind provision of his nonlinear force-free extrapolation code for validation runs related to this work. MKG acknowledges fruitful discussions with M. A. Berger, A. A. Pevtsov, and A. Nindos, as well as invaluable mentoring by D. M. Rust and the late B. J. LaBonte, whose inspiration was the driving force behind a series of studies on solar magnetic helicity. We also gratefully acknowledge the Institute of Space Applications and Remote Sensing (ISARS) of the National Observatory of Athens for the availability of their computing cluster facility for runs related to this work. Hinode is a Japanese mission developed and launched by ISAS/JAXA, with NAOJ as domestic partner and STFC (UK) as international partners. It is operated by these agencies in co-operation with ESA and NSC (Norway). This work has received partial support from NASA's Guest Investigator Grant NNX08AJ10G and from the European Union's Seventh Framework Programme (FP7/2007-2013) under grant agreement $n^o$ PIRG07-GA-
2010-268245. Finally, we thank the anonymous referee who has contributed substantially to the improvement of the analysis described in this work.  
\appendix
\section{Calculation of possible $\mcal{L}_{ac}^{arch}$-values for two discrete magnetic flux tubes $a$ and $c$}
Per \citet{demoulin_etal_06}, the mutual-helicity parameter $\mcal{L}_{ac}=\mcal{L}_{ac}^{close}+\mcal{L}_{ac}^{arch}$ of a pair of flux tubes $a$ and $c$ can be calculated by progressively bringing the interacting pair from infinity to its prescribed position and geometry. This can be generalized for any set of discrete flux tubes, where $(a,c)$ is now any given pair of tubes belonging to the set. For each case of interaction there are only {\it two} possible values of 
$\mcal{L}_{ac}^{arch}$ that depend on (i) the geometry of the pair, as reflected by the line segments formed by the footpoint locations of each tube, and (ii) whether tube $a$ is ``above'' tube $c$ ($\mcal{L}_{ac}^{arch} \equiv \mcal{L}_{\hat{a}c}^{arch}$) or tube $c$ is above tube $a$ ($\mcal{L}_{ac}^{arch} \equiv \mcal{L}_{a\hat{c}}^{arch}$). By stating that a tube is ``above'' another tube we imply that its apex is higher than the appex of its mate, with respect to the anchoring boundary that defines the open volume. \citet{demoulin_etal_06} further demonstrated a practical way of calculating $\mcal{L}_{ac}^{arch}$, namely by means of interior angles of the triangles formed by footpoint segments on the anchoring boundary plane. In the following we calculate $\mcal{L}_{ac}^{arch}$ for all cases pertinent to our study, both reproducing the values of \citet{demoulin_etal_06}, and deriving values for cases that were not examined by these authors. 

In practice, $\mcal{L}_{ac}^{arch}$ is the mean angle by which each line segment "sees'' the other, normalized over $\pi$. Therefore, assuming a flux tube $a$ with positive- and negative-polarity footpoints $a_+$ and $a_-$, respectively (so its footpoint segment is $a_+ a_-$), each footpoint ``sees'' the dipole $c$ (with footpoint segment $c_+ c_-$) by angles $\alpha _{a_+}$ and $\alpha _{a_-}$, respectively. Then, in case $a$ is ``above'' $c$ we have 
\beq
\mcal{L}_{\hat{a}c}^{arch} = {{1} \over {2 \pi}} (\alpha _{a_+} + \alpha _{a_-})\;\;.
\label{Lac_a}
\eeq
Similarly, footpoints $c_+$ and $c_-$ of tube $c$ will ``see'' the segment $a_+ a_-$ with angles $\alpha _{c_+}$ and $\alpha _{c_-}$, respectively. In case $c$ is ``above'' $a$ we have 
\beq
\mcal{L}_{a\hat{c}}^{arch} = {{1} \over {2 \pi}} (\alpha _{c_+} + \alpha _{c_-})\;\;.
\label{Lac_c}
\eeq
It should be mentioned that interior angles $\alpha_{x_{\pm}}$ 
($x \equiv \{ a, c \}$) do not necessarily follow the trigonometric (counterclockwise) convention, so they can be positive or negative. Moreover, it is clear from Equations (\ref{Lac_a}) and (\ref{Lac_c}) that 
$|\mcal{L}_{ac}^{arch}| < 1$.

All possible footpoint-segment configurations in our calculations appear in Figure \ref{fs}. Cases of non-intersecting (Figure \ref{fs}a) and intersecting (Figure \ref{fs}b) segments were also examined by \citet{demoulin_etal_06}. Given that our connectivity matrix has been constructed by connecting magnetic partitions, however, it is very common to find multiple connections connecting a given partition with others. Since all connections are viewed as slender flux tubes originating from the partition's flux-weighted centroid (Figures \ref{mtx_fig}a, \ref{mtx_fig}b) we have cases of ``matching'' footpoints, as well (Figures \ref{fs}c, \ref{fs}d). The apparent conflict with the principle that magnetic field lines cannot intersect may be raised by clarifying that these ``matching'' footpoints are, in fact, distinct footpoints but with distances falling into unresolved length scales within a given partition. Therefore, for the sake of simplicity they are thought to originate from the same well-known location, 
i.e., the partition's flux-weighted centroid. This introduces some modifications in $\mcal{L}_{ac}^{arch}$-values and we calculate these modifications here. 

The four possible locations of footpoint-segment geometry, as illustrated in Figure \ref{fs}, are as follows: 

\ben
\item[CASE A:] Non-intersecting footpoints (Figure \ref{fs}a). It can then be found that the footpoints of flux tube $a$ ``see'' flux tube $c$ by the angles 
$$
\alpha _{a_+} = \theta _{c_- a_+} - \theta _{c_+ a_+}\;\;\;\;\;and\;\;\;\;\;
\alpha _{a_-} = \theta _{c_+ a_-} - \theta _{c_- a_-}\;\;,
$$
where $\theta _{x_{\pm} y_{\pm}}$ ($x,y \equiv \{ a,c \}$) are the azimuth angles of the segments $y_{\pm}x_{\pm}$ (that is, with the trigonometric-circle origin at 
$x_{\pm}$). Correspondingly, the footpoints of flux tube $c$ ``see'' flux tube $a$ by the angles 
$$
\alpha _{c_+} = -\theta _{c_+ a_+} + \theta _{c_+ a_-}\;\;\;\;\;and\;\;\;\;\;
\alpha _{c_-} = -\theta _{c_- a_-} + \theta _{c_- a_+}\;\;.
$$ 
Obviously, then, from equations (\ref{Lac_a}), (\ref{Lac_c}) we obtain
\beq
\mcal{L}_{\hat{a}c}^{arch} = \mcal{L}_{a \hat{c}}^{arch}\;\;,
\label{case_a}
\eeq
reproducing \citet{demoulin_etal_06}. In essence, in case of non-intersecting footpoint segments the two possible $\mcal{L}_{ac}^{arch}$-values collapse to a single value, regardless of appex heights for the two flux tubes. 
\item[CASE B:] Intersecting footpoints (Figure \ref{fs}b). In case $a$ is ``above'' $c$ (upper configuration) we find 
$$
\alpha _{a_+} = -\theta _{c_+ a_+} + \theta _{c_- a_+}\;\;\;\;\;and\;\;\;\;\;
\alpha _{a_-} = 2 \pi - \theta _{c_- a_-} + \theta _{c_+ a_-}\;\;.
$$
In case $c$ is ``above'' $a$ (lower configuration), we further have 
$$
\alpha _{c_+} = -\theta _{c_+ a_+} + \theta _{c_+ a_-}\;\;\;\;\;and\;\;\;\;\;
\alpha _{c_-} = -\theta _{c_- a_-} + \theta _{c_- a_+}\;\;, 
$$
From Equations (\ref{Lac_a}) and (\ref{Lac_c}), then, we find 
\beq
\mcal{L}_{\hat{a}c}^{arch} - \mcal{L}_{a\hat{c}}^{arch} =1\;\;,
\label{case_b3}
\eeq
also reproducing \citet{demoulin_etal_06}.
\item[CASE C:] Positive ``matching'' footpoints (Figure \ref{fs}c). In this case we obviously have 
$$
\alpha _{a_+} = \alpha _{c_+} = 0\;\;.
$$
To calculate the remaining interior angles $\alpha _{a_-}$ and $\alpha _{c_-}$ we first notice that the formed triangle dictates 
\beq
|\phi _{ac}| + |\alpha _{a_-}| + |\alpha _{c_-}|= \pi\;\;.
\label{c_1}
\eeq
The orientation (sign) of $\alpha _{a_-}$, $\alpha _{c_-}$ depends on the orientation of the angle $\phi _{ac}$ between segments $a$ and $c$. Angle $\alpha _{a_-}$ {\it always} follows the sign of $\phi _{ac}$ while angle $\alpha _{c_-}$ {\it always} shows opposite orientation. Depending on the orientation of the triangle as a rigid shape, one may further find from trigonometric analysis that the magnitudes of $\alpha _{a_-}$, $\alpha _{c_-}$ are given by 
\begin{eqnarray}
\begin{array}{lll}
|\alpha _{a_-}| & = & |\theta _{c_-a_-} - \theta _{a_-a_+} \pm \pi|\\
|\alpha _{c_-}| & = & \left\{
\begin{array}{l}
|\theta _{c_-a_+} - \theta _{c_-a_-}|\\
|\theta _{c_-a_+} - \theta _{c_-a_-} \mp 2 \pi|\;\;.\\
\end{array}
\right.
\end{array}
\label{c_2}
\end{eqnarray}
It then becomes trivial to find which of the six possible cases applies for 
$|\alpha _{a_-}|$, $|\alpha _{c_-}|$ so that Equations (\ref{c_2}) satisfy 
Equation (\ref{c_1}). The orientation of $\alpha _{a_-}$, $\alpha _{c_-}$ is then found when the respective orientation of $\phi _{ac}$ is found, that is, by determining whether one forms an interior (hence $< \pi$) angle from $a$ to $c$ following (+) or opposing (-) the trigonometric convention. For the example of Figure \ref{fs}c we obviously have 
$\phi _{ac} >0$, so $\alpha _{a_-} >0$ and $\alpha _{c_-} <0$. 

Calculating the $\mcal{L}_{ac}^{arch}$-values in this case we {\it always} find 
\beq
\mcal{L}_{\hat{a}c}^{arch} \mcal{L}_{a \hat{c}}^{arch} <0\;\;,
\label{c_3}
\eeq
which makes it straightforward to determine the preferred value so that 
$\Delta E_{c_{mut}} >0$ in Equation (\ref{DEc_mut2}). 
\item[CASE D:] Negative ``matching'' footpoints (Figure \ref{fs}d). In this case we have 
$$
\alpha _{a_-} = \alpha _{c_-} = 0\;\;.
$$
The analysis is symmetric to that of Case C in the sense that $\alpha _{a_+}$ always follows the orientation of $\phi _{ac}$ and $\alpha _{c_+}$ always opposes it. From the formed triangle we have 
\beq
|\phi _{ac}| + |\alpha _{a_+}| + |\alpha _{c_+}|= \pi\;\;, 
\label{d_1}
\eeq
and a trigonometric analysis for different triangle orientations shows that 
\begin{eqnarray}
\begin{array}{lll}
|\alpha _{a_+}| & = & |\theta _{c_+a_+} - \theta _{a_+a_-} \pm \pi|\\
|\alpha _{c_-}| & = & \left\{
\begin{array}{l}
|\theta _{c_+a_-} - \theta _{c_+a_+}|\\
|\theta _{c_+a_-} - \theta _{c_+a_+} \mp 2 \pi|\;\;.\\
\end{array}
\right.
\end{array}
\label{d_2}
\end{eqnarray}
We solve the systems of Equations (\ref{d_1}), (\ref{d_2}) to determine 
$|\alpha _{a_+}|$, $|\alpha _{c_+}|$ and we infer the orientation of $\phi _{ac}$ to determine their orientations. In this case, as well, we find 
\beq
\mcal{L}_{\hat{a}c}^{arch} \mcal{L}_{a \hat{c}}^{arch} <0\;\;,
\label{d_3}
\eeq
so determining the desired value for $\Delta E_{c_{mut}} >0$ in Equation (\ref{DEc_mut2}) is straightforward. 
\een
\section{Uncertainties for Magnetic Free Energy and Relative Magnetic Helicity}
Assuming that the uncertainty in the calculation of the potential magnetic energy $E_p$ is negligible, the uncertainty $\delta E_t$ of the total magnetic energy equals the uncertainty $\delta E_c$ of the free magnetic energy $E_c$. Hence 
\beq
\delta E_c = \delta E_t\;\;.
\label{ap1}
\eeq
As the free magnetic energy comprises of self and mutual terms, $E_{c_{self}}$ and $E_{c_{mut}}$, respectively, the uncertainty $\delta E_c$ depends on the uncertainties of these two terms, i.e.
\beq 
\delta E_c = \sqrt{(\delta E_{c_{self}})^2 + (\delta E_{c_{mut}})^2}\;\;,
\label{ap2}
\eeq
assuming, of course, that $E_{c_{self}}$ and $E_{c_{mut}}$ are independent of each other. 

Likewise, the uncertainty $\delta H_m$ of the relative magnetic helicity depends on the respective uncertainties of the self and mutual helicity terms, i.e.
\beq 
\delta H_m = \sqrt{(\delta H_{m_{self}})^2 + (\delta H_{m_{mut}})^2}\;\;.
\label{ap3}
\eeq
\subsection{Uncertainties in the self terms of free energy and relative helicity}
From Equation (\ref{Ec_single}), the uncertainty $\delta E_{c_{(s)}}$ in case of a single force-free flux tube (that is, without mutual terms) depends on the uncertainties $\delta \alpha$, $\delta A$, and $\delta \lambda$ of the force-free parameter $\alpha$ of the tube and the fitting parameters $A$ and $\lambda$, respectively, of Equation (\ref{srel}), assuming that the tube's flux content $\Phi$ is known without uncertainty. By standard error propagation and assuming that $\alpha$, $A$, and $\lambda$ are independent of each other, we find 
\beq
\delta E_{c_{(s)}} = E_{c_{(s)}} [ 4 ({{\delta \alpha} \over {\alpha}})^2 + 
                                ({{\delta A} \over {A}})^2 + 
                              4 (\ln \Phi)^2 (\delta \lambda)^2 ]^{1/2}\;\;.
\label{ap4}
\eeq
From Equation (\ref{ap4}) and assuming a collection of $N$ flux tubes, the uncertainty $\delta E_{c_{self}}$ in the self term of the free magnetic energy (Equation (\ref{Ec_self})) will then be
\beq
\delta E_{c_{self}} = \sqrt{\sum _{f=1}^N (\delta E_{c_{(s)},f})^2}\;\;.
\label{ap5}
\eeq

Likewise, for a single force-free flux tube the relative magnetic helicity is given by Equation (\ref{Hm_single}). Therefore, the uncertainty $\delta H_{m_{(s)}}$ in this case is 
\beq
\delta H_{m_{(s)}} = |H_{m_{(s)}}| [ ({{\delta \alpha} \over {\alpha}})^2 +
                                ({{\delta A} \over {A}})^2 +
                                4 (\ln \Phi)^2 (\delta \lambda)^2 ]^{1/2}\;\;.
\label{ap6}
\eeq
Then, from Equation (\ref{Hm_self}) in case of a collection of $N$ flux tubes, the uncertainty $\delta H_{m_{self}}$ in the self term of the relative magnetic helicity is 
\beq
\delta H_{m_{self}} = \sqrt{\sum _{f=1}^N (\delta H_{m_{(s)},f})^2}\;\;.
\label{ap7}
\eeq

In both Equations (\ref{ap5}) and (\ref{ap7}) each uncertainty $\delta \alpha _f$
of the involved flux tubes' force-free parameters is calculated by Equation (\ref{da_ij}), thus involving the uncertainties of the flux-weighted $\alpha$-values of the connected partitions. 
\subsection{Uncertainties in the mutual terms of free energy and relative helicity}
From Equation (\ref{Ec_mut}), the mutual term of the free energy for a given pair $(l,m)$ of flux tubes is given by 
\beq
E_{c_{mut},lm} = {{1} \over {8 \pi}} \alpha _{lm} \mcal{L}_{lm}^{arch} \Phi _l \Phi _m\;\;.
\label{ap8}
\eeq
The uncertainty $\delta E_{c_{mut},lm}$ is then given by 
\beq 
\delta E_{c_{mut},lm} = E_{c_{mut},lm} \sqrt{ 
                     ({{\delta \alpha _{lm}} \over {\alpha _{lm}}})^2 + 
                     ({{\delta \mcal{L}_{lm}^{arch}} \over  
                       {\mcal{L}_{lm}^{arch}}})^2}\;\;,
\label{ap9}
\eeq
ignoring dependencies between $\alpha _{lm}$ and $\mcal{L}_{lm}^{arch}$ for simplicity. Now assuming that each mutual-energy term $E_{c_{mut},lm}$ is independent from the other terms, the overall uncertainty $\delta E_{c_{mut}}$ in the mutual term of the free magnetic energy is 
\beq
\delta E_{c_{mut}} = \sqrt{ \sum _{l=1}^N \sum _{m=1, l \ne m}^N 
(\delta E_{c_{mut},lm})^2 }\;\;, 
\label{ap10}
\eeq 
for a collection of $N$ flux tubes. 
Therefore, the problem becomes equivalent to determining the uncertainties 
$\delta \alpha _{lm}$, where $\alpha _{lm}=(1/2)(\alpha _l + \alpha _m)$, and $\delta \mcal{L}_{lm}^{arch}$, for each mutual-energy term (Equation (\ref{ap9})). 

For $\delta \alpha _{lm}$ we trivially have 
\beq 
\delta \alpha _{lm} = {{1} \over {2}} \sqrt{ (\delta \alpha _l)^2 + 
                                             (\delta \alpha _m)^2}\;\;, 
\label{ap11}
\eeq
assuming that $\alpha _l$ and $\alpha _m$ are independent of each other. The uncertainty $\delta \mcal{L}_{lm}^{arch}$ of $\mcal{L}_{lm}^{arch}$ depends sensitively on the uncertainty $\delta \alpha _{lm}$. In particular: 
\bit
\item[$\bullet$] In case one of the two possible $\Delta E_{c_{mut}}$-values in Equation (\ref{DEc_mut2}) is positive but the sign of $\alpha _{lm}$ is ambiguous ($\delta \alpha _{lm} \ge |\alpha _{lm}|$), any of the two possible $\mcal{L}_{lm}^{arch}$-values could be used. In case the sign of $\alpha _{lm}$ is known with certainty ($\delta \alpha _{lm} < |\alpha _{lm}|$) then the choice of one of the two possible $\mcal{L}_{lm}^{arch}$-values is definitive. Therefore, 
\begin{eqnarray}
\delta \mcal{L}_{lm}^{arch} & = & \left\{
\begin{array}{lll}
   |\mcal{L}_{\hat{l}m}^{arch} - \mcal{L}_{l\hat{m}}^{arch}| &;& if\;\;
\Delta E_{c_{mut}}^{(1)} \Delta E_{c_{mut}}^{(2)} <0\;\;\;and\;\;\;\delta \alpha _{lm} \ge |\alpha _{lm}|\\  
   0 &;& if\;\;\Delta E_{c_{mut}}^{(1)} \Delta E_{c_{mut}}^{(2)} <0\;\;\;and\;\;\;\delta \alpha _{lm} < |\alpha _{lm}|\;\;.\\
\end{array}
\right.
\label{ap12}
\end{eqnarray}
\item[$\bullet$] In case both possible $\Delta E_{c_{mut}}$-values in Equation (\ref{DEc_mut2}) are either positive or negative but the sign of $\alpha _{lm}$ is ambiguous ($\delta \alpha _{lm} \ge |\alpha _{lm}|$), then either the selected $\mcal{L}_{lm}^{arch}$-value or $\mcal{L}_{lm}^{arch}=0$ could be used. In case 
$\delta \alpha _{lm} < |\alpha _{lm}|$, the choice of the non-zero or zero $\mcal{L}_{lm}^{arch}$ is definitive. Because this situation can only occur in case footpoint segments are not intersecting, so 
$\mcal{L}_{\hat{l}m}^{arch}=\mcal{L}_{l\hat{m}}^{arch}$, we define an uncertainty 
\begin{eqnarray}
\delta \mcal{L}_{lm}^{arch} & = & \left\{
\begin{array}{lll}
   |\mcal{L}_{lm}^{arch}| &;& if\;\;
\Delta E_{c_{mut}}^{(1)} \Delta E_{c_{mut}}^{(2)} >0\;\;\;and\;\;\;\delta \alpha _{lm} \ge |\alpha _{lm}|\\  
   0 &;& if\;\;\Delta E_{c_{mut}}^{(1)} \Delta E_{c_{mut}}^{(2)} >0\;\;\;and\;\;\;\delta \alpha _{lm} < |\alpha _{lm}|\;\;.\\
\end{array}
\right.
\label{ap13}
\end{eqnarray}
\eit

When the mutual term of the relative helicity is concerned, for a given pair 
$(l,m)$ of flux tubes we have 
\beq
H_{m_{mut},lm} = \mcal{L}_{lm}^{arch} \Phi _l \Phi _m\;\;.
\label{ap14}
\eeq
Assuming, as usual, that $\Phi _l$, $\Phi _m$ are known without uncertainties, the 
uncertainty $\delta H_{m_{mut},lm}$ is given by 
\beq
\delta H_{m_{mut},lm} = |H_{m_{mut},lm} | \delta \mcal{L}_{lm}^{arch}\;\;.
\label{ap15}
\eeq
Then, the overall uncertainty $\delta H_{m_{mut},lm}$ is trivially given by 
\beq
\delta H_{m_{mut}} = \sqrt { \sum _{l=1}^N \sum _{m=1, l \ne m}^N 
                            (\delta H_{m_{mut},lm})^2 }\;\;,
\label{ap16}
\eeq
where each uncertainty $\delta H_{m_{mut},lm}$ is calculated by Equation (\ref{ap15}).
%
\bibliography{ms_refs}

\clearpage
\newpage
\centerline{\includegraphics[width=7.cm,angle=0]{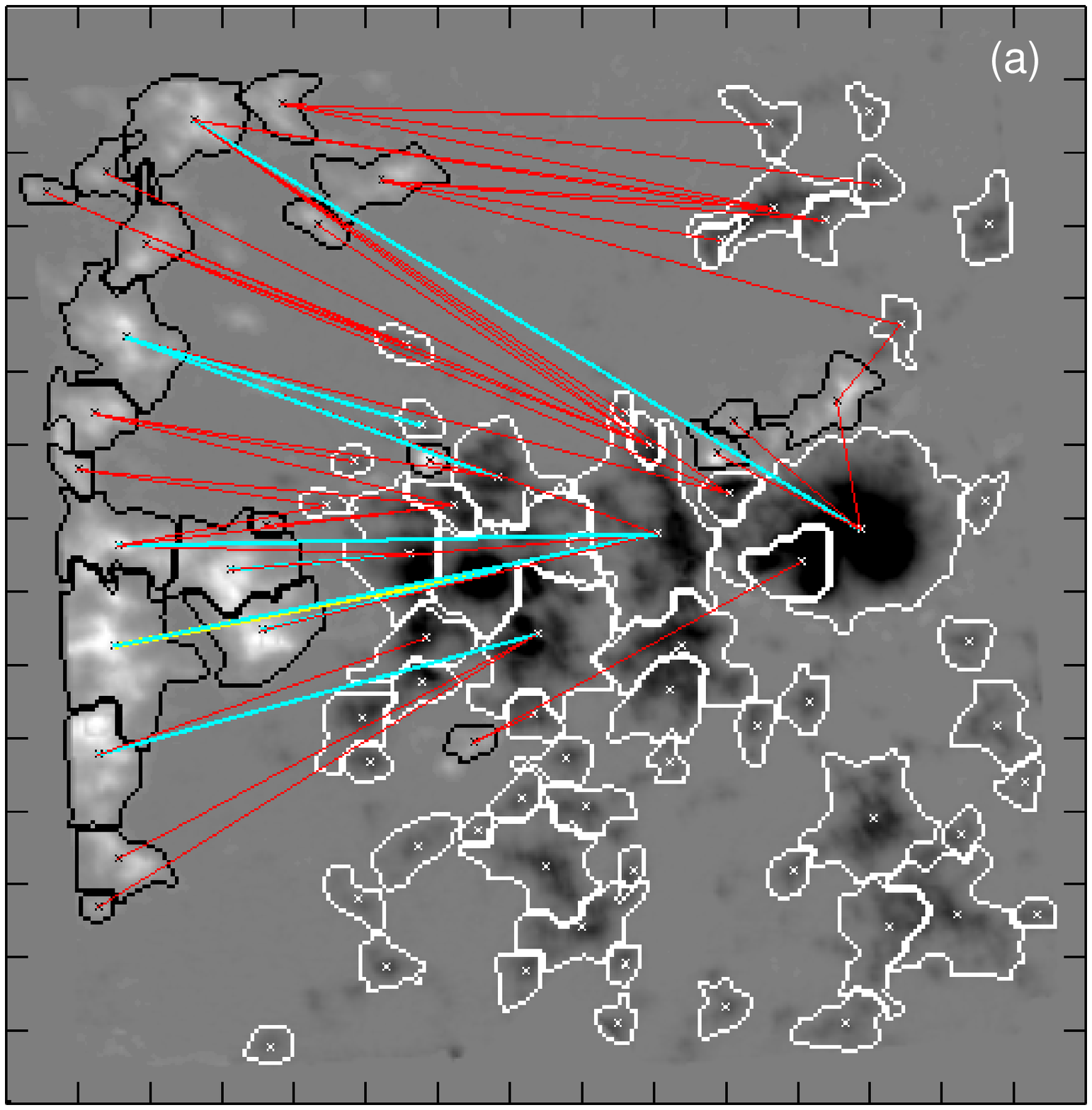}
            \includegraphics[width=7.cm,angle=0]{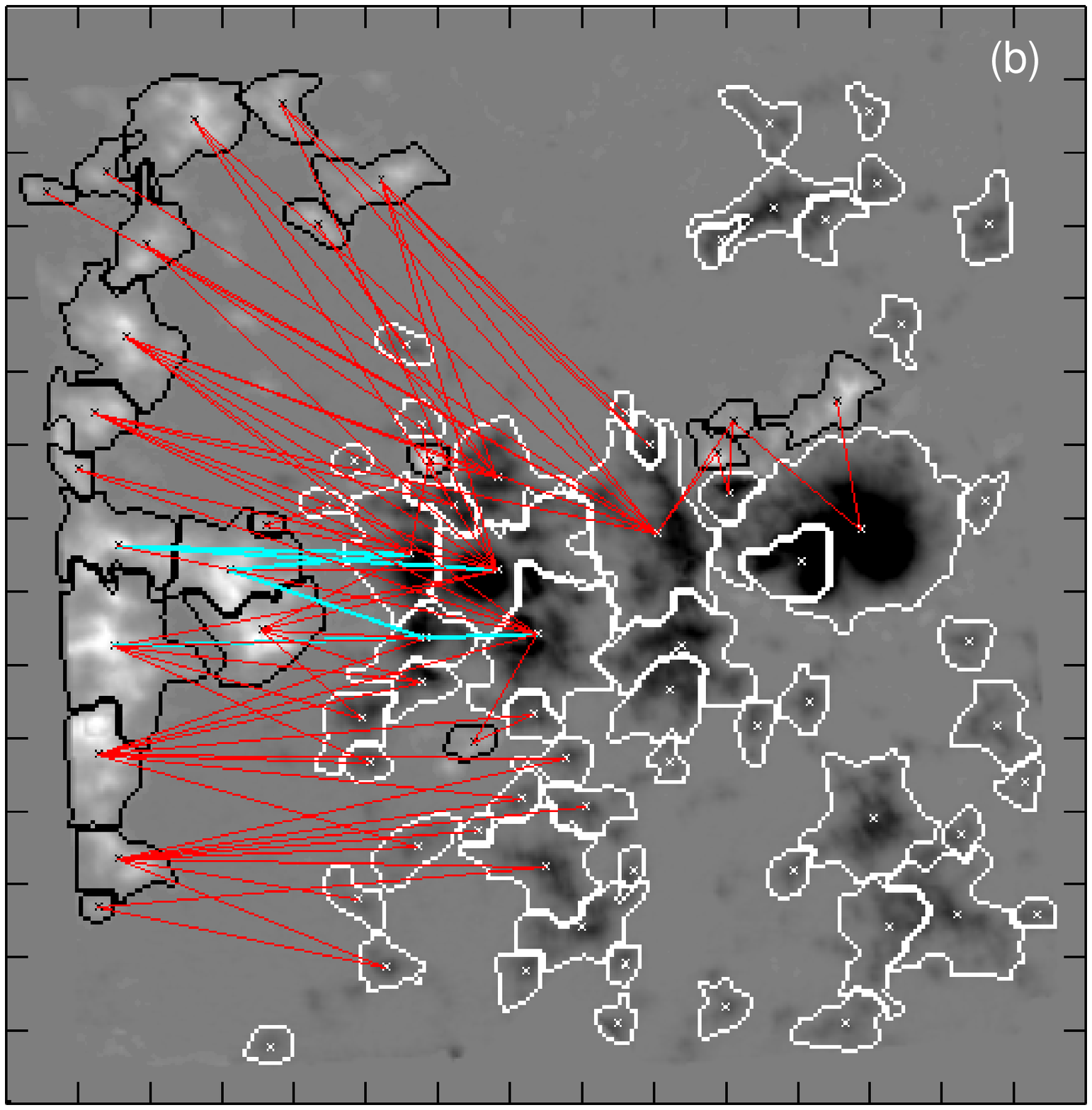}}
\centerline{\includegraphics[width=8.5cm,angle=0]{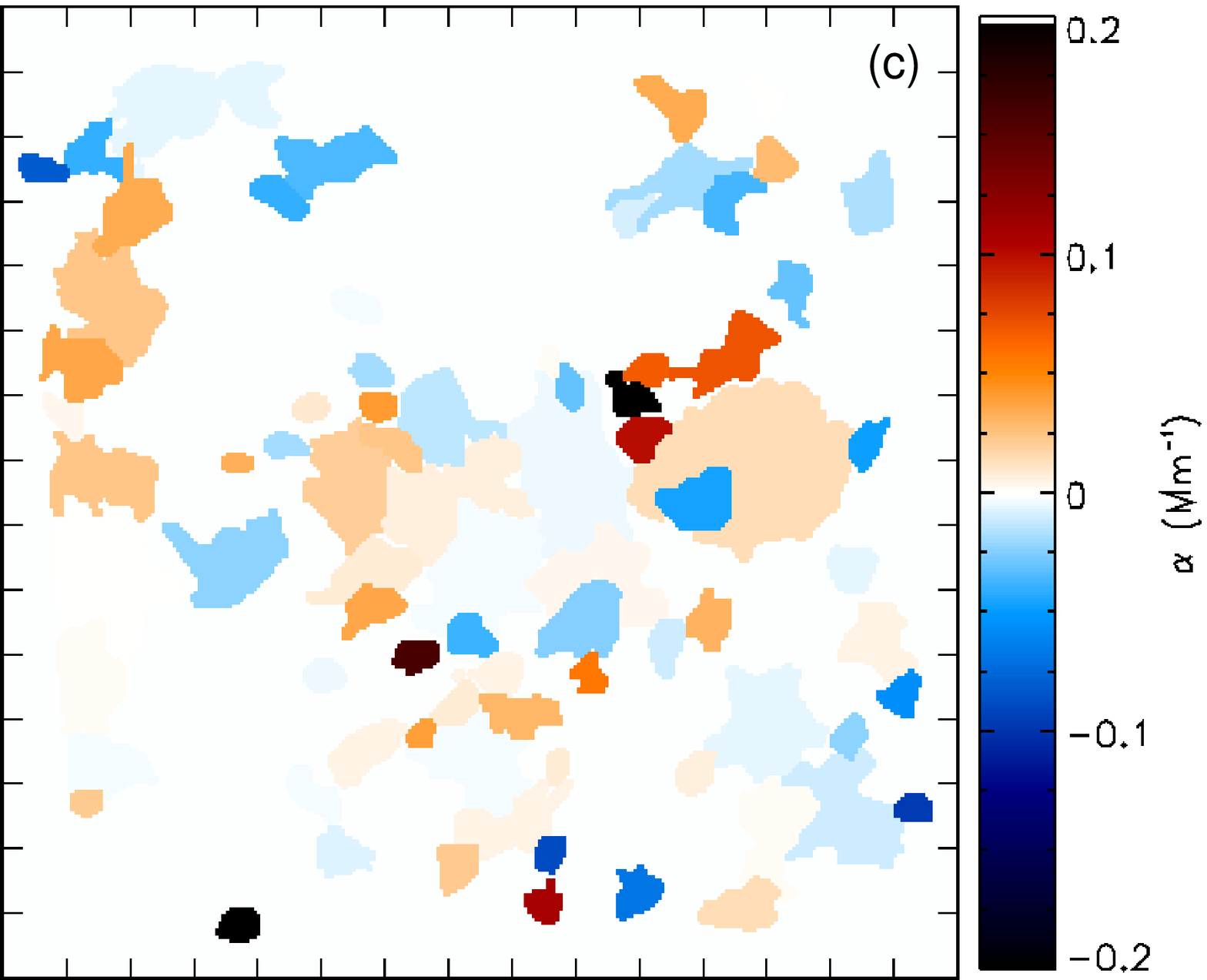}}
\figcaption{Example magnetic 
  connectivity in NOAA AR 10254, observed by the IVM on
  2003 January 13, at 21:44 UT. Images (a) and (b) show the vertical
  magnetic field component  
  in grayscale, saturated at $\pm 1000\;G$ with the contours
  bounding the identified magnetic partitions. Image (c) shows the
  respective flux-weighted $\alpha$-value for each partition. The 
  flux tubes identified by the magnetic connectivity matrices are represented
  by line segments of different colors and thicknesses, connecting the
  flux-weighted centroids of the respective pair of 
  partitions. Only closed connections within the field of view are
  shown and considered; disconnected partitions are exclusively linked to 
  opposite fluxes beyond the field of view. Red, cyan, and yellow
  segments denote magnetic flux contents within the ranges  
  $[5 \times 10^{19}, 5 \times 10^{20}]\;Mx$, 
  $[5 \times 10^{20}, 10^{21}]\;Mx$, and $[10^{21}, 5 \times 10^{21}]\;Mx$,
  respectively. The connectivity matrices shown enclose (a) $\sim 10^{22}$ Mx for 
  the simulated-annealing and (b) $\sim 8.3 \times 10^{21}$ Mx for the   
  potential-field connectivity. Tic mark separation is $20\arcsec$. 
  North is up; west is to the right. 
\label{mtx_fig}}
\newpage
\centerline{\includegraphics[width=17.cm,angle=0]{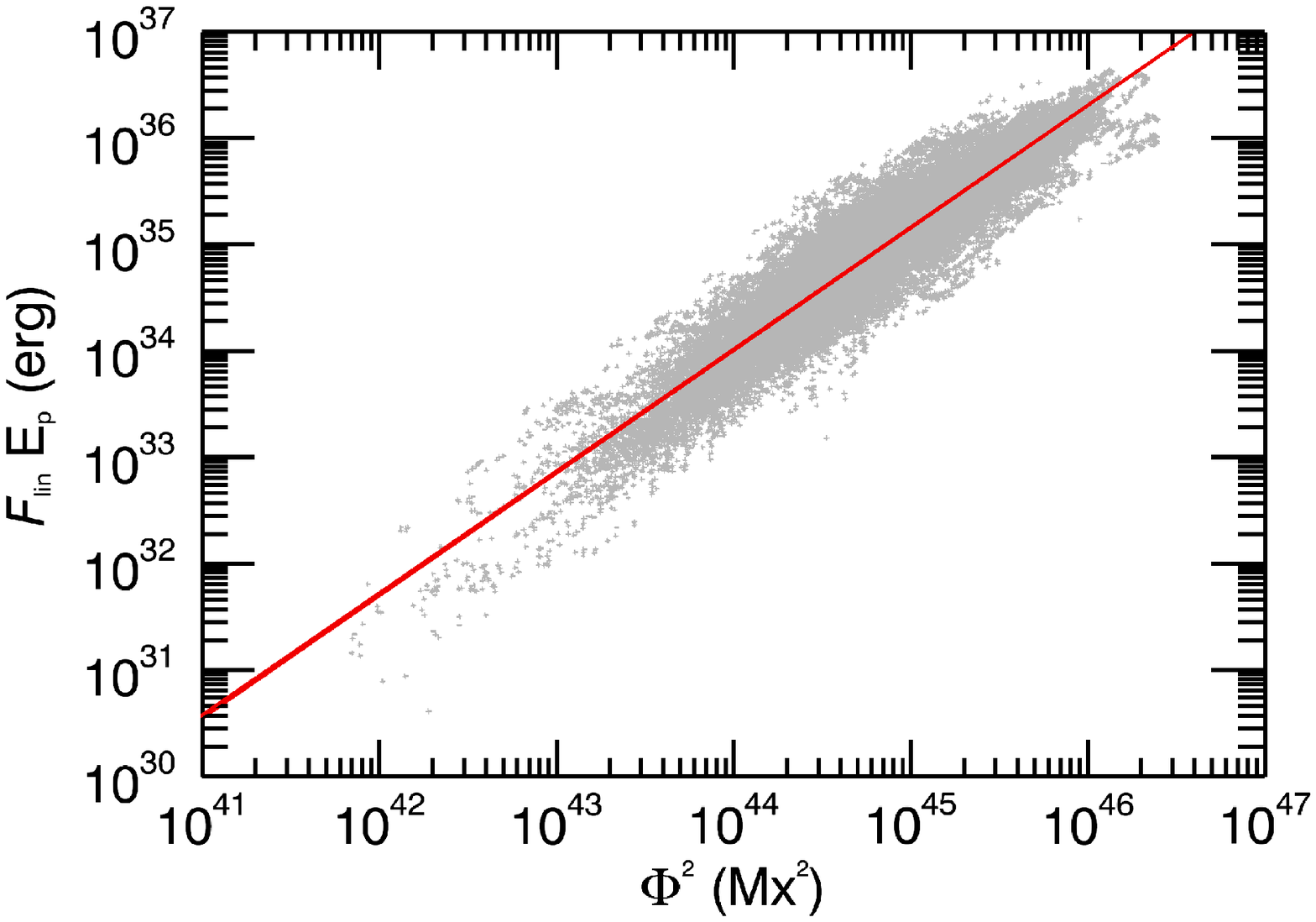}}
\figcaption{Relation between the "scaled'' potential energy 
  $\mcal{F}_{lin}E_p$ and the
  square $\Phi ^2$ of the unsigned magnetic flux for 56686
  active-region line-of-sight magnetograms recorded by SoHO/MDI
  between 1996 and 2005. All these regions are located within 
  $\pm 30^o$ from the solar central meridian at the time of
  observation. The red line is the least-squares best fit described by 
  Equation (\ref{srel}).
\label{scat_plot}}
\newpage
\centerline{\includegraphics[width=11.cm,angle=0]{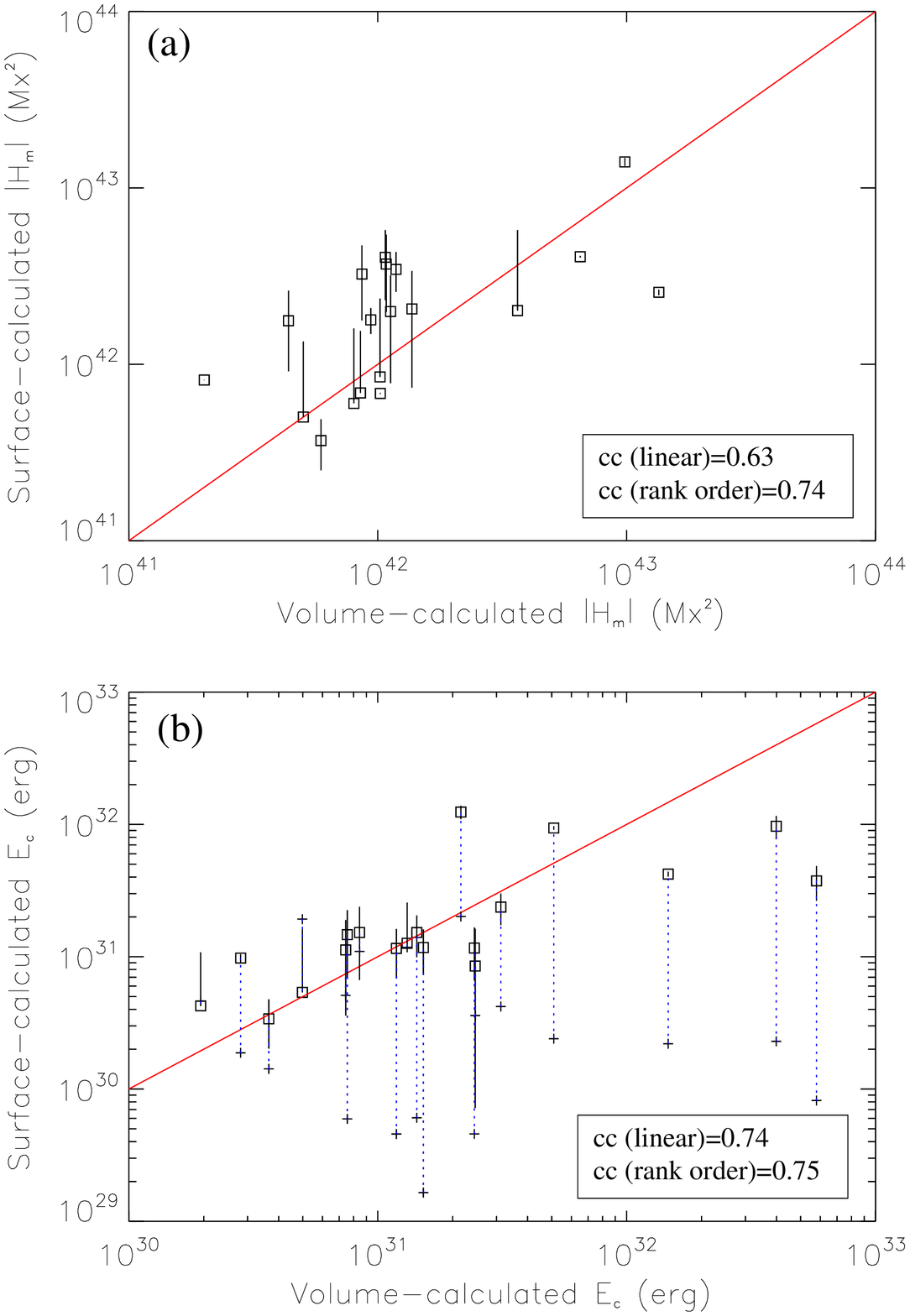}}
\figcaption{Comparison between our NLFF surface-calculated values (ordinate) and well-known volume-calculated values (abscissa) for (a) the relative magnetic helicity ($|H_m|$) and (b) the free magnetic energy ($E_c$) budgets. Equality between the two budgets is denoted by the red lines. In (b), crosses denote the Woltjer-Taylor free magnetic energies $E_{c_{WT}}$ (Equation (\ref{Ec_wt})) linked to the respective free magnetic energies $E_c$ by  blue dotted lines. Error bars (only to higher values in some cases, as including the lower-value error bars would result in negatives in these cases) correspond to our NLFF surface-calculated values for both plots.  
\label{val_plot}}
\newpage
\centerline{\includegraphics[width=11.cm,angle=0]{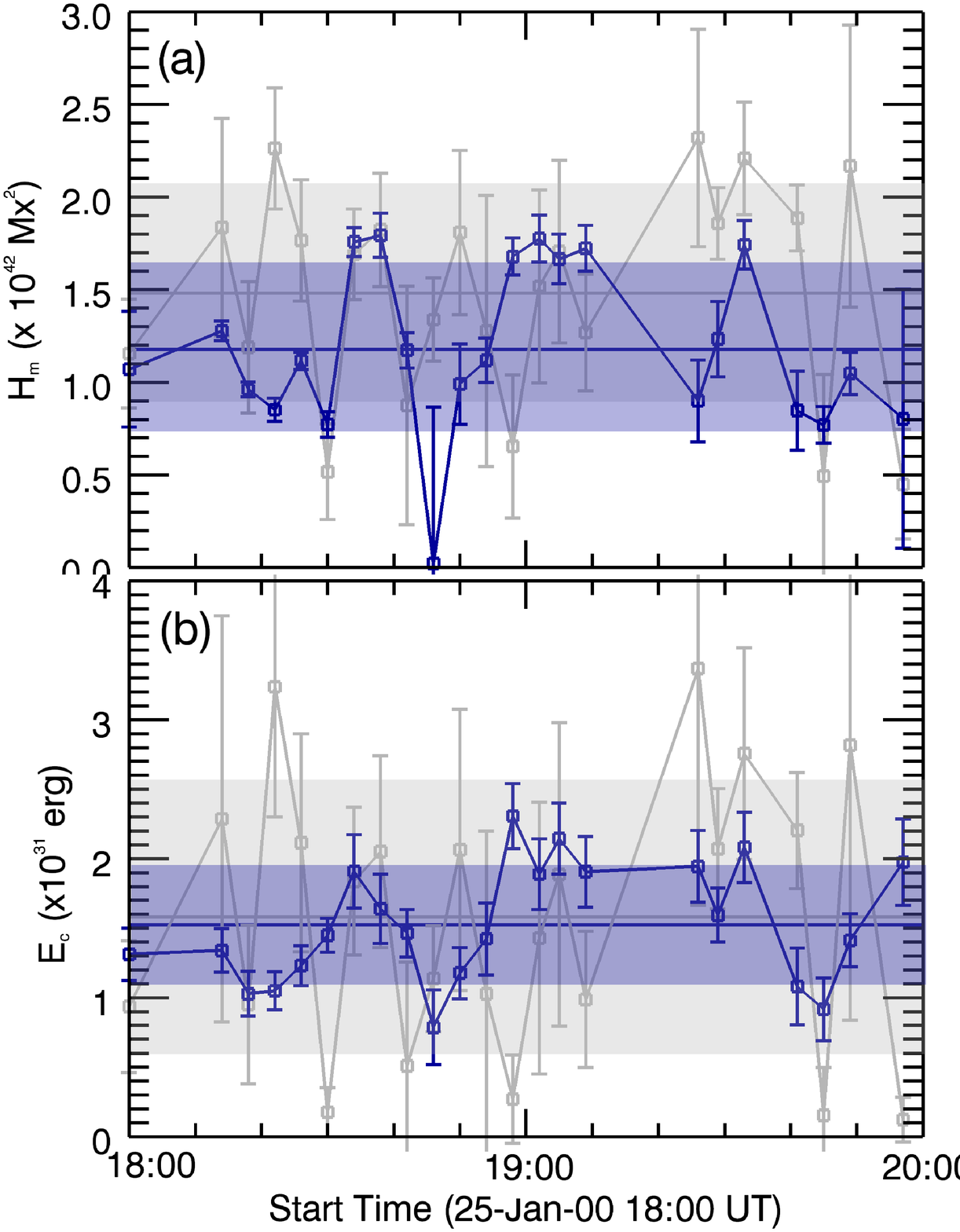}}
\figcaption{Comparison between the LFF and NLFF field approximations in calculating (a) the relative magnetic helicity budget $H_m$ and (b) the free magnetic energy budget $E_c$ in NOAA AR 8844, observed by the IVM for nearly two hours on 2000 January 25. The mean values $\bar{H}_m$ and $\bar{E}_c$ are indicated by straight lines (blue for the NLFF-, gray for the LFF-field calculations) while the respective standard deviations are indicated by the blue- and gray-shaded areas. 
\label{8844_comp}}
\newpage
\centerline{\includegraphics[width=11.cm,angle=0]{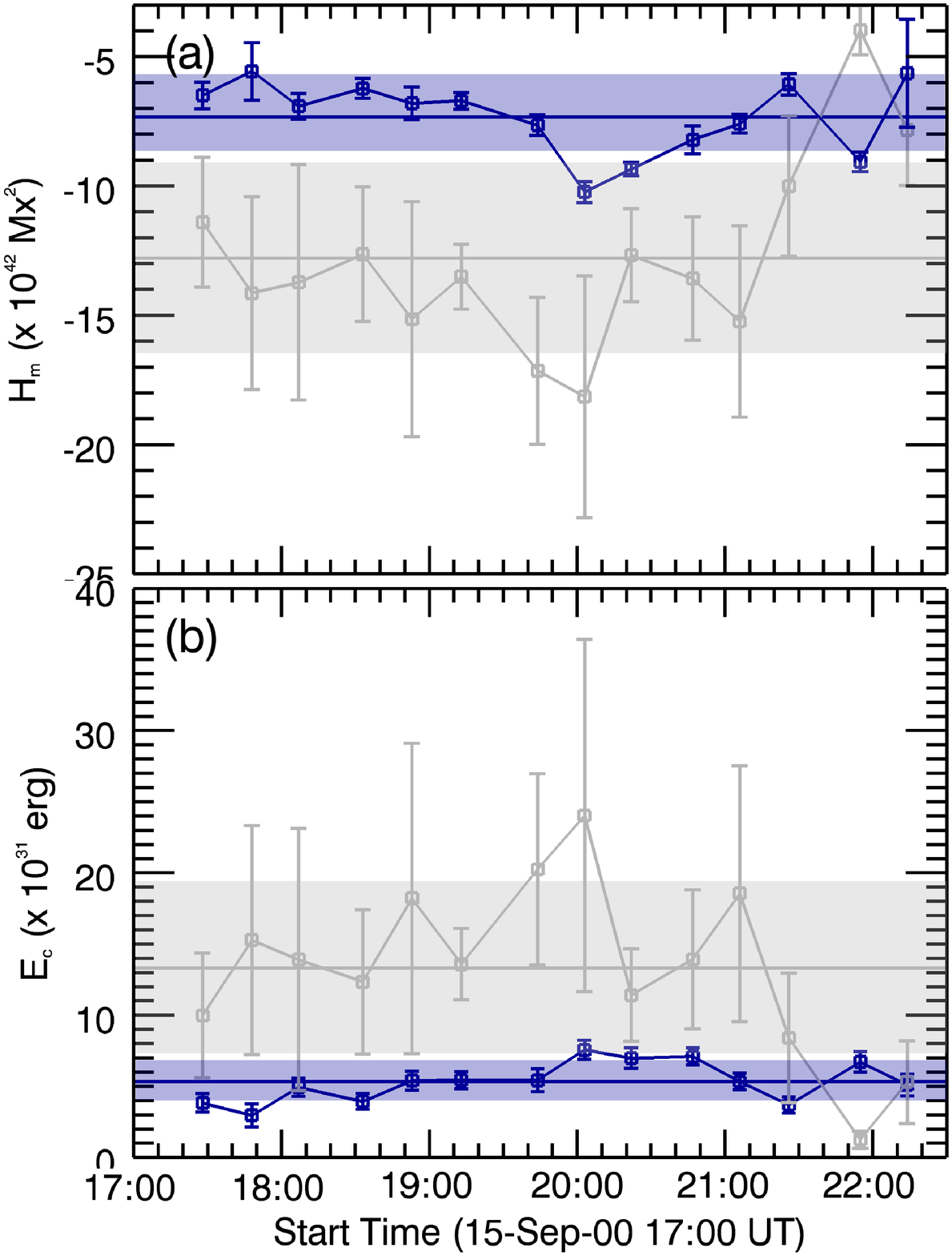}}
\figcaption{Same as in Figure \ref{8844_comp} but for NOAA AR 9165, observed by the IVM for nearly five hours on 2000 September 15. 
\label{9165_comp}}
\newpage
\centerline{\includegraphics[width=17.cm,angle=0]{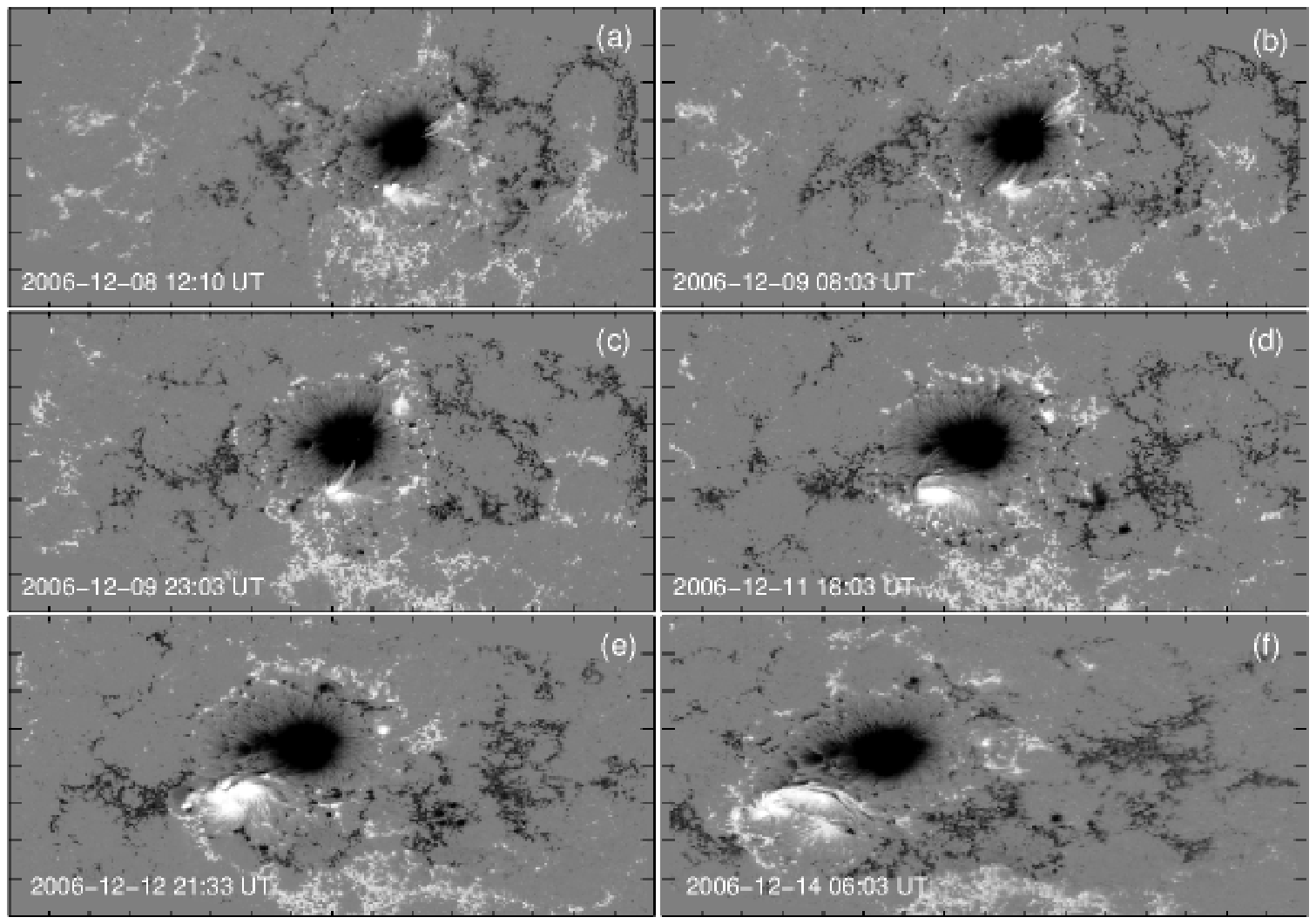}}
\figcaption{Heliographic vertical field components of selected vector SOT/SP magnetograms of NOAA AR 10930 binned to $\sim 0.62\arcsec$ per pixel. The dates and universal times of the start of observations for each magnetogram are also shown. The vertical field component is saturated at $\pm 2000$ G.Tic mark separation is $20\arcsec$. North is up; west is to the right. 
\label{10930_im}}
\newpage
\centerline{\includegraphics[width=10.cm,angle=0]{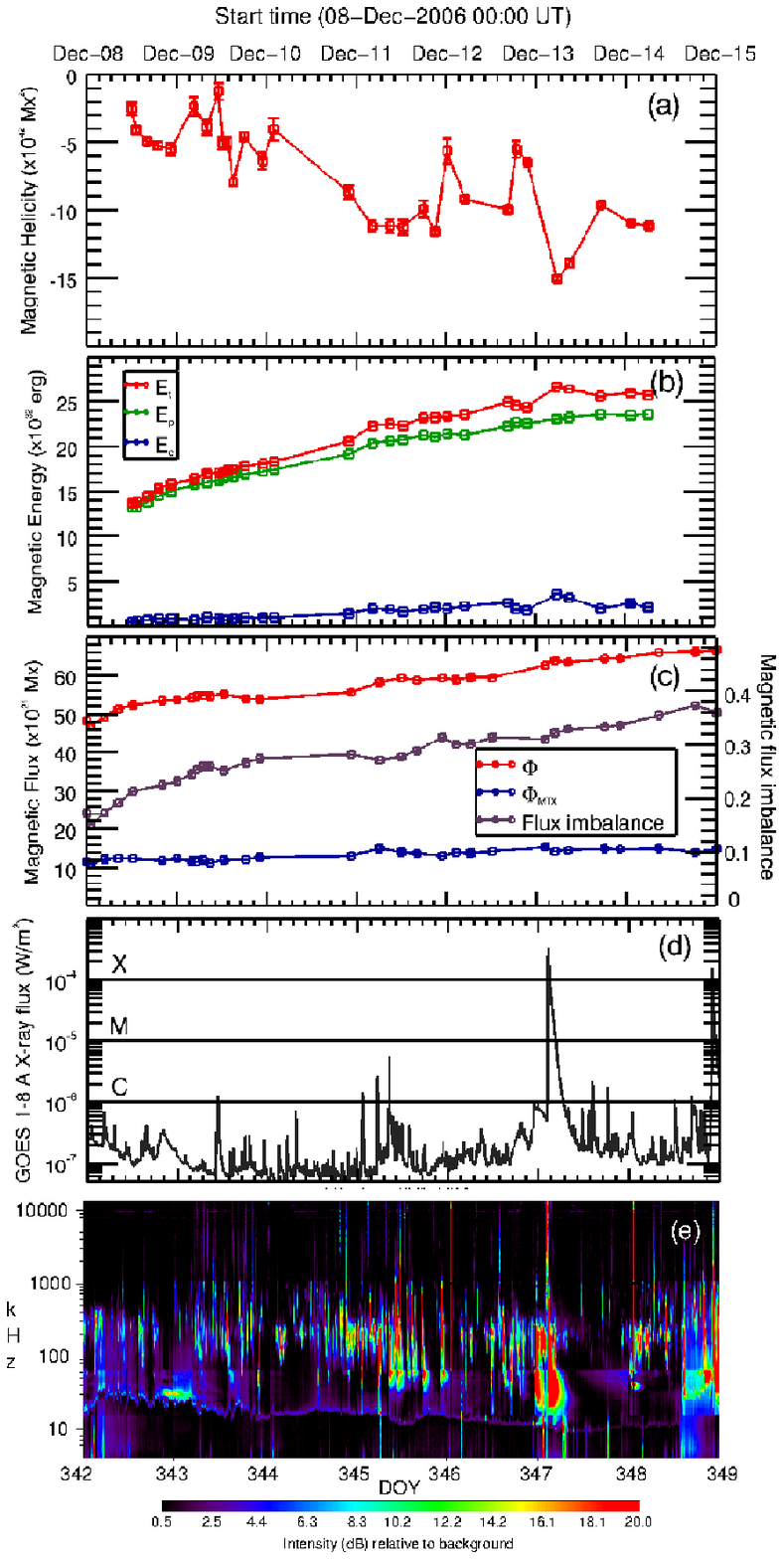}}
\figcaption{Temporal evolution of NOAA AR 10930, observed over a period of a few days in 2006 December by Hinode's SOT/SP, as reflected on (a) the calculated relative magnetic helicity budget, (b) the calculated magnetic energy budgets (with the total, potential-field, and free, magnetic energy shown by red, green, and blue curves, respectively), and (c) the unsigned magnetic flux (red) and flux imbalance (purple), including the unsigned flux participating in the connectivity matrix (blue). Also shown for reference are the respective (d) GOES 1-8 \AA$\;$ solar X-ray flux, showing flaring activity, and (e) WIND/WAVES frequency-time radio spectrum, with Type-II activity indicating shock-fronted CMEs. All flares and CMEs originate from the region because it was the only AR present in the solar disk at the time of interest. 
\label{10930_calc}}
\newpage
\centerline{\includegraphics[width=13.cm,angle=0]{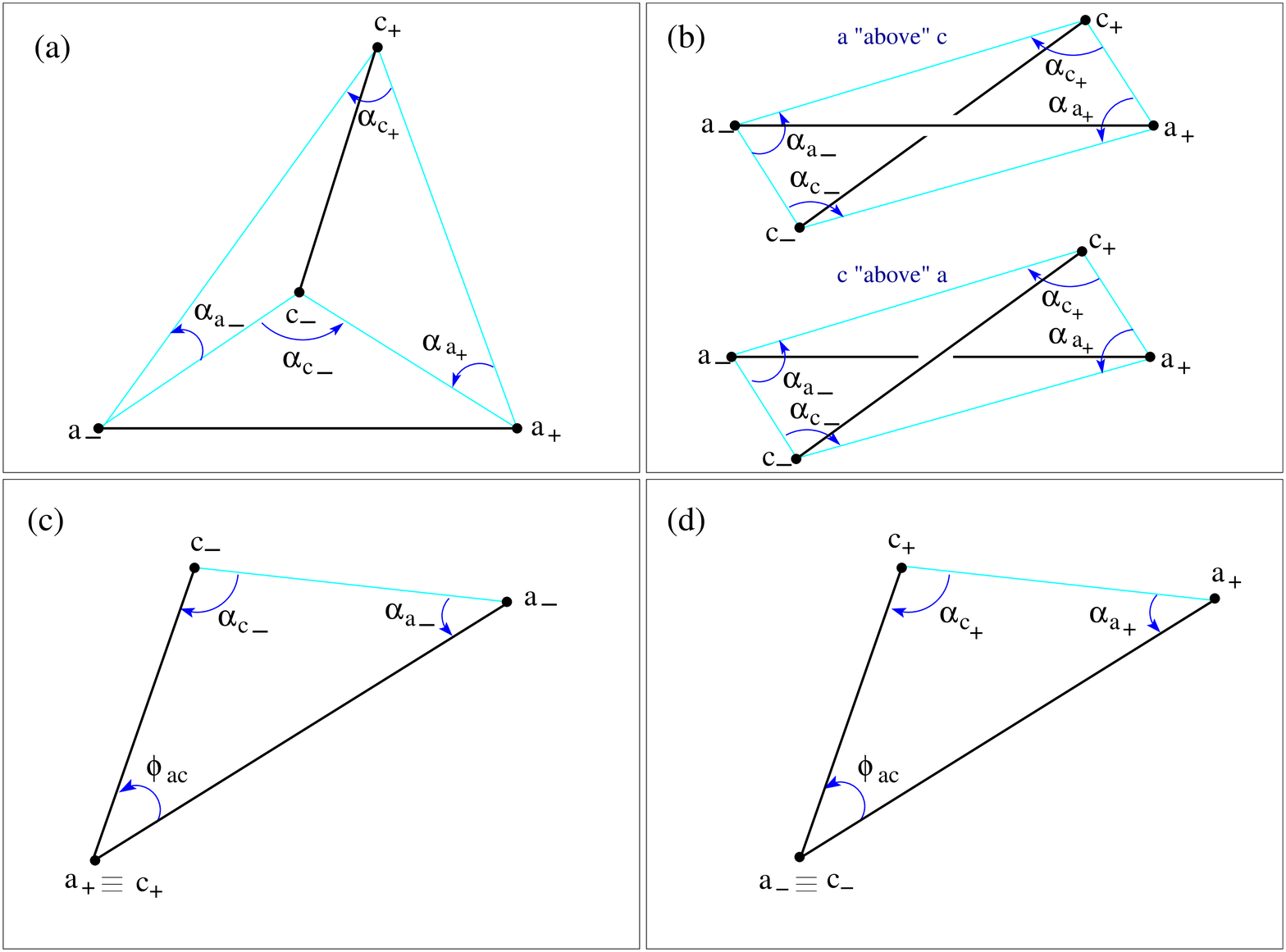}}
\figcaption{Possible geometrical positions of the photospheric footpoints of two discrete, arched flux tubes $a$ and $c$: (a) non-intersecting footpoint segments, (b) intersecting footpoint segments, including cases where tube $a$ is ``above'' tube $c$ (upper sketch) and  where tube $c$ is ``above'' tube $a$ (lower sketch), (c) ``matching'' positive-polarity footpoints, and (d) ``matching'' negative-polarity footpoints. In all sketches, footpoint segments are depicted by thick black lines, while all connecting lines are cyan. All interior angles of the formed triangles are also shown. Cases (a) and (b) were first studied by \citet{demoulin_etal_06} but they are reproduced here for reasons of completeness. 
\label{fs}}
\newpage
\begin{table}
\begin{tabular}{lccccc}
\hline
\hline
NOAA AR & $\bar{\Phi}$  & $\bar{\Phi}_{MTX}$   & $\bar{E}_p$ & 
$\bar{E}_c$ & $\bar{H}_m$\\
        & ($\times 10^{21}$ Mx) & ($\times 10^{21}$ Mx) & ($\times 10^{32}$ erg) &
($\times 10^{32}$ erg) & ($\times 10^{42}\; Mx^2$)\\
\hline
8844...... & $5.1 \pm 0.2$ & $3.6 \pm 0.2$ & $2.97 \pm 0.01$ & $0.152 \pm 0.04$ & 
$1.18 \pm 0.5$\\
9165...... & $17.1 \pm 0.8$ & $12.8 \pm 0.4$ & $9.52 \pm 0.5$ & $0.53 \pm 0.14$ & 
$-7.3 \pm 1.4$\\
Ratio ...... & $3.4 \pm 0.2$ & $3.6 \pm 0.2$ & $3.2 \pm 0.2$ & $3.5 \pm 1.3$ & 
$6.2 \pm 2.9$\\
\hline
\hline
\end{tabular}
\caption{Synopsis of the mean unsigned magnetic flux ($\bar{\Phi}$), connected flux (i.e. unsigned flux participating in the magnetic connectivity matrix; $\bar{\Phi}_{MTX}$), current-free magnetic energy ($\bar{E}_p$), free magnetic energy ($\bar{E}_c$), and relative magnetic helicity ($\bar{H}_m$) budgets for NOAA ARs 8844 and 9165 as obtained by our NLFF methodology. The third row
refers to the ratio $|P_{9165}/P_{8844}|$ between a given parameter
$P_{9165}$ of NOAA AR 9165 and the respective parameter $P_{8844}$
of NOAA AR 8844.
In terms of mean free magnetic energy $\bar{E}_c$, $\sim 99.9$\% for NOAA AR 8844 and $\sim 99.5$\% for NOAA AR 9165 are contributed by the mutual terms. For the mean relative magnetic helicity $\bar{H}_m$  
the mutual-term contributions are $\sim 99.9$\% for NOAA AR 8844 and 
$\sim 99.6$\% for NOAA AR 9165.}
\label{Tb1}
\end{table} 
\end{document}